\documentclass[manuscript,prb,aps,superscriptaddress,floatfix]{revtex4} 
\usepackage{graphicx,amssymb,amsmath}

\begin{document}

\title{A Molecular Dynamics Computer Simulation Study of Room-Temperature Ionic Liquids:\\
I.\ Equilibrium Solvation Structure and Free Energetics}

\author{Y. Shim} 
\affiliation{Department of Physics, Seoul National University, Seoul 151-747, Korea} 
\affiliation{Department of Chemistry, Carnegie Mellon University, Pittsburgh, PA 15213, U.S.A.} 

\author{M. Y. Choi} 
\affiliation{Department of Physics, Seoul National University, Seoul 151-747, Korea} 
\affiliation{Korea Institute for Advanced Study, Seoul 130-722, Korea} 

\author{Hyung J. Kim}
\altaffiliation{Author to whom correspondence should be addressed. E-mail: hjkim@cmu.edu} 
\affiliation{Department of Chemistry, Carnegie Mellon University, Pittsburgh, PA 15213, U.S.A.} 

\begin{abstract}
Solvation in 1-ethyl-3-methylmidazolium chloride and in 1-ethyl-3-methylimidazolium hexafluorophosphate  near equilibrium is investigated via molecular dynamics computer simulations with diatomic and benzenelike molecules employed as probe solutes.  It is found that electrostriction plays an important role in both solvation structure and free energetics.  The angular and radial distributions of cations and anions become more structured and their densities near the solute become enhanced as the solute charge separation 
grows.  Due to the enhancement in structural rigidity induced by electrostriction, the force constant associated with solvent configuration fluctuations relevant to charge shift and transfer processes is also found to increase.  The effective polarity and reorganization free energies of these ionic liquids are analyzed and compared with those of highly polar acetonitrile. Their screening behavior of electric charges is also investigated.
\end{abstract}


\maketitle

\section{Introduction}
Molecular ionic liquids based on bulky cations such as N,\,N$^{\prime}$-dialkylimidazolium and N-alkylpyridinium have been the subject of intensive experimental study.\cite{ref1}  This class of systems can exist as a liquid at or near room temperature and is resistant to water.  Their vapor pressure is essentially zero, so that volatile products can be separated completely through distillation and the ionic liquids can be recycled.  Accordingly, they provide an environmentally benign (``green'') alternative to toxic organic solvents, which has potentially a wide range of applications in organic synthesis and separation chemistry.\cite{ref1}

Structure and solvation properties of these room-temperature ionic liquids (RTILs) have been studied via a variety of spectroscopic methods.\cite{ref8,ref2,bright,pandey,ref3,ref4,kerr:exp,gordon,maroncelli:phosphonium,petrich}
Many solvatochromic measurements show that the effective polarity of RTILs is comparable to that of highly dipolar solvents such as acetonitrile and small alcohols.  Another interesting finding is that dynamic Stokes shifts in imidazolium-based RTILs determined with the picosecond time resolution are considerably smaller than the corresponding static shifts.\cite{ref3,ref4,comment:phosphonium} 
Though somewhat controversial,\cite{petrich} this suggests that a substantial part of solvent relaxation in RTILs occur on a subpicosecond time scale despite their high viscosity.\cite{ref3,ref4} 

There have been growing efforts to study RTILs via molecular dynamics (MD) simulation methods.\cite{ref7,ref:margulis,ref9,stassen,shim,lynden-bell:benzene,kobrak,lopes,ribeiro}  Earlier works have focused mainly on pure liquids, such as structure and transport properties of RTILs.\cite{ref7,ref:margulis,ref9,stassen,lopes,ribeiro}  Recently, we examined equilibrium solvation properties of 1-ethyl-3-methylimidazolium chloride ($\mathrm {EMI^+Cl^-}$) and 1-ethyl-3-methylimidazolium hexafluorophosphate ($\mathrm{EMI^+PF_6^-}$) in the presence of a model diatomic solute.\cite{shim}   We found that  in accord with experiments,\cite{ref2,bright,pandey} the effective polarity of these RTILs is very high and nearly comparable to that of ambient water.  Solvent fluctuation dynamics are characterized by at least two totally different time scales: rapid subpicosecond dynamics followed by slow non-exponential relaxation due to ion transport.\cite{shim}  Furthermore, the fast subpicosecond dynamics account for more than 50~\% of the entire relaxation of solvent fluctuations.  Within the context of linear response, this finding lends strong support to the existence of ultrafast nonequilibrium solvent relaxation in RTILs based on the indirect experimental evidence.\cite{ref3,ref4} 

In this article and its companion\cite{rtil:shim2} -- hereafter, referred to as II -- we extend our previous MD study\cite{shim} to investigate solvation structure and dynamics in RTILs both in and out of equilibrium.  By employing diatomic and benzenelike solutes as probe molecules, we make a detailed analysis of the solvent structural variations with the solute charge distributions and their influence on solvation free energetics.  We also examine free energy curves along a solvent coordinate relevant to charge shift and transfer processes in solution and quantify outer-sphere reorganization free energy.  Time-dependent Stokes shifts subsequent to an instantaneous change in the solute charge distribution are analyzed and comparison is made with equilibrium solvent fluctuation dynamics.  

The outline of this paper is as follows: In Sec.~II we give a brief account of the model description and simulation methods employed in our study.  The MD results for solvation structure and free energy fluctuations in EMI$^+$Cl$^-$ and in EMI$^+$PF$_6^-$ are presented in Sec.~III, while Sec.~IV concludes. Dynamic solvation properties are studied in II.

\section{Model Description and Simulation Methods}

The MD simulations are performed using the DL$\_$POLY program.\cite{ref5}
The simulation cell is comprised of a single rigid solute molecule immersed in either EMI$^+$Cl$^-$ or EMI$^+$PF$_6^-$ solvent, consisting of 112 pairs of rigid cations and anions. 
Two different types of solutes are considered: a diatomic molecule and a benzenelike one (see Table~\ref{table1}).  Two constituent atoms of the first type are fixed at a separation of 3.5\,\AA\ and interact with the solvent through Lennard-Jones (LJ) and Coulombic interaction potentials.
The LJ parameters, $\sigma\!=\!4$\,\AA\ and $\epsilon/k_{\rm B}\!=\!100$\,K with Boltzmann's constant $k_{\rm B}$, and mass, $m\!=\!100$\,amu, are chosen to be identical for each constituent atom of the diatomic solute molecule.\cite{comment:solute:parameter} 
Three different charge distributions are considered, i.e., a neutral pair (NP) with no charges, an ion pair (IP) with unit charge separation ($q\!=\!\pm e$), and their intermediate state  with $q\!=\!\pm 0.5 e$ referred to as HIP, where $e$ is the elementary charge.  As for the benzenelike solute, we employ $\sigma\!=\!3.5$\,\AA\ and $\epsilon/k_{\rm B}\!=\!40.3$\,K for carbon and $\sigma\!=\!2.5$\,\AA\ and $\epsilon/k_{\rm B}\!=\!25.2$\,K  for hydrogen.\cite{benzene}  In addition to the axially symmetric charge distribution appropriate for the regular ground-state benzene (RB), we also study another charge distribution, referred to as ``dipolar'' benzene (DB); this differs from RB by $\pm e$ in charge assignments for two C atoms in the para positions while the remaining four carbon and six hydrogen atoms have the same partial charges as in the RB solute.  This yields the dipole moment difference of 13.5\,D between the RB and DB charge distributions.

The solvent parameters employed are the same as in our prior MD study.\cite{shim} To be specific, we used the AMBER force field\cite{ref6} 
for the LJ parameters and the partial charge assignments of Ref.~\onlinecite{ref7} for EMI$^+$.  The united atom representation was employed for the CH$_2$ and CH$_3$ moieties of the ethyl group of the cation as well as for the methyl group  (Fig.~\ref{fig:emi}).  Experimental geometry determined by X-ray diffraction\cite{ref8} was used.  As for Cl$^-$, we used $\sigma\!=\!4.4$\,\AA\ and $\epsilon/k_{\rm B}\!=\!50.4$\,K.
PF$_6^-$ was described as a united atom with $\sigma\!=\!5.6$\,\AA\ and $\epsilon/k_{\rm B}\!=\!200$\,K.  The numerical values for the solute and solvent parameters employed in the present study are compiled in Table~\ref{table1}.

The simulations are conducted in the canonical ensemble using the Nos\'e-Hoover extended system method.\cite{ref11}   We study two different densities at temperature $T=400$\,K for each RTIL solvent: $\rho=1.1$ and 1.2\,g/cm$^{3}$ for EMI$^+$Cl$^-$ and $\rho=1.31$ and 1.375\,g/cm$^{3}$ for EMI$^+$PF$_6^-$.    Periodic, cubic boundary conditions are employed. Long-range electrostatic interactions are computed via the Ewald method with a conducting surrounding medium, resulting in essentially no truncation of these interactions. 
The trajectories are integrated via the Verlet leapfrog algorithm combined with the quaternion method for rotations with a time step of 2\,fs. 
Equilibrium simulations are carried out with 2\,ns equilibration, followed by a 4\,ns trajectory from which averages are computed.

\section{Results and Discussions}

The MD results for EMI$^+$Cl$^-$ and EMI$^+$PF$_6^-$ are summarized in Table~\ref{table:result} and Figs.\ \ref{fig2}--\ref{fig:free}.  We begin our analysis by considering the equilibrium solvent structure around the solutes.

\subsection{Equilibrium Solvation Structure}
\label{subsec:structure}

In Fig.~\ref{fig2} radial distribution functions $g(r)$ of the solvent cations and anions around the diatomic solutes in EMI$^+$Cl$^-$ are displayed.  For convenience, the midpoint of two nitrogen atoms of the imidazolium ring is employed as the cation center in the present study.  The most salient feature in Fig.~\ref{fig2} is that the solvent structure varies markedly with the solute charge distribution.   
As the magnitude of the charges at both sites of the solute is increased, their respective first solvation shells consisting mainly of the solvent ions of the opposite charge become more pronounced and localized.  For example, the Cl$^-$ distribution around the positive site of IP in Fig.~\ref{fig2}(a) vanishes almost completely between its first and second peaks, while it shows no substantive structure in the presence of NP.  This reveals the development of a well-defined shell of Cl$^-$ around the IP solute.  As the extent of the solute charge separation increases, the position of the first peak in $g(r)$ moves in and its height grows.  This results in the solvent density increase close to the solute: For instance, the number of Cl$^{-}$ in the first solvation shell of the solute (+) site, i.e., those located within 5~\AA\ from the site, increases from 1.8 for NP to 3.1 and 4.1 for HIP and IP, respectively (see also Fig.~\ref{fig3ad} below).  
We refer to this structure-making in RTILs with increasing solute charge separation and accompanying local solvent density enhancement as electrostriction. 
Similar electrostrictive behaviors of dipolar solvents play an important role in various solvation and reaction processes involving charge shift and transfer.\cite{kohnstam,ladanyi,honig,berne,kim:cav} 
The basic characteristics of the cation distributions in Figs.\ \ref{fig2}(b) and~\ref{fig2}(c) are essentially the same as those of the anion distributions in Fig.~\ref{fig2}(a).  We notice that the cation distributions are somewhat less structured than the chloride distribution. This is probably due to the extended structure and charge distribution of EMI$^+$ compared with Cl$^{-}$.  

For additional insight into electrostriction, we consider the integrated ion number  $N(r)$
\begin{equation}
N(r)=4\pi n \int_0^r g(r')\, {r'}^2 dr'\ ,
\label{eq:int:density}
\end{equation}
which measures the number of anions (or cations) located within a distance $r$ from the solute.  Here $n$ is the number density of anions (or cations) and $g(r)$ is their radial distribution around the solute atom sites.   The results for Cl$^{-}$ and EMI$^+$ center in the presence of the NP and IP solutes are presented in Fig.~\ref{fig3ad}.  Compared with the NP case, $N(r)$ for Cl$^{-}$ around the (+) site of IP is enhanced considerably up to $r\approx 6$\,\AA, while that around the ($-$) site is somewhat reduced.  The qualitative behavior of EMI$^+$ center in Fig.~\ref{fig3ad}(b) is similar to that of anions.  The strong electrostatic interactions between the IP solute charges and solvent ions are responsible for the $N(r)$ trend in the vicinity of the solute.
We, however, notice that this trend becomes reversed if we consider a larger region around the solute.  To be specific, the total number of Cl$^{-}$ present inside a sphere of radius $r\approx 6$--9\,\AA\ around the (+) site of IP is smaller than that inside the same volume but in the presence of the NP solute.  By contrast, the total number of Cl$^{-}$ inside a similar sphere 
centered at the $(-)$ site of IP exceeds that in the presence of NP.   We attribute the former (latter) to the presence of excess negative (positive) chages close to the solute positive (negative) site, which repel (attract) Cl$^{-}$ ions in the outer region (cf.\ Fig.~\ref{fig:screen_DP} below).  To see this better, we consider a sphere of radius, say 4\,\AA, centered at the (+) site of IP (cf.\ Fig.~\ref{fig2}(a)).  If we calculate the total charge inside this spherical volume including the solute charges, we obtain $-3.2 e$.\cite{comment:local:charge}$^{a}$  The positive solute charge at the center attracts Cl$^{-}$ strongly, so that a small neighborhood surrounding the (+) site becomes highly negatively-charged.  As a consequence, the immdediate vicinity of the negatively-charged region is not easily accessible by Cl$^{-}$ ions from the outer region and thus becomes anion-deficient and cation-rich.  This is clearly demonstrated in Fig.~\ref{fig2}(a), where the chloride radial distribution around the (+) site of IP nearly vanishes between $r\approx 4$\,\AA\  and 6\,\AA.
If we consider a larger volume around the solute (+) site to include the above-mentioned cation-rich and anion-deficient region, we find that the total charge enclosed inside changes the sign and becomes positive; for example, with a radius of 7\,\AA,  the total charge becomes $+3.2e$.\cite{comment:local:charge}$^{b}$  This will then attract Cl$^{-}$ and repel EMI$^{+}$ to create an anion-rich and cation-deficient region right outside.  
This shows that the solute charges exert a strong influence on solvent structure by inducing anion-rich and cation-deficient, and anion-deficient and cation-rich regions that alternate in the solvent medium.  The oscillatory character of the solvent charge distributions around the solute is studied in detail in Fig.~\ref{fig:screen_DP} below.

The cation and anion distributions with respect to the center of the diatomic solutes in EMI$^+$Cl$^-$ are exhibited in Fig.~\ref{fig3}.  It is noteworthy that $g(r)$ in Fig.~\ref{fig3}(a) is considerably less structured than that in Fig.~\ref{fig2}(a).  This indicates that the first solvation shell of anions is formed mainly around the positive site of the solute (see below).  Though much lesser in extent, the cations show a similar trend. 
While the structural variations of the first solvation shells are most dramatic, the modulations of the solvent radial distributions by the solute charges are not limited to the short range.
$g(r)$ in Figs.\ \ref{fig2} and~\ref{fig3} and $N(r)$ in Fig.~\ref{fig3ad} vary substantially with the solute electronic structure for $r\sim 10$\,\AA.  This implies that the influence of the solute charges on the solvent ions extends well beyond a separation of $\sim\!10$\,\AA. 
We will come back to this point below. 


Figure~\ref{fig4} shows the angular probability distribution $P(\cos\theta)$ of EMI$^+$ and of Cl$^-$ around the diatomic solute, where $\theta$ denotes the angle between the solute dipole vector and the solute-to-ion center-to-center direction.
(For example, $\theta = 0^\circ$ and 180$^\circ$ correspond to the directions of the positive and negative sites of the solute from its center, respectively.)
  The ions are grouped together in three different regions according to the distance $r$ from the solute center: region~I ($r< 5.5$\,\AA), region~II ($5.5<r<9$\,\AA) and region~III ($r>9$\,\AA).  In each region, the probability is normalized to unity. 
In the presence of the NP solute, $P(\cos\theta)$ for both cations and anions in regions II and~III are found to be independent of $\theta$; i.e.,  their distributions are isotropic.  In region~I, on the other hand, while nearly constant around $\cos\theta=0$, $P(\cos\theta)$ vanishes as $\cos\theta$ approaches $\pm 1$.  This anisotropy arises from the presence of the solute diatoms at $\theta=0^\circ$ and 180$^\circ$.  Because of their extended structure, the cation distribution in region~I  around $\cos\theta=0$ is not as isotropic as the corresponding anion distribution.  

The angular distributions of the ions around the IP solute in Figs.\ \ref{fig4}(c) and~\ref{fig4}(d) differ markedly from those  in Figs.\ \ref{fig4}(a) and~\ref{fig4}(b).  The former are much more structured than the latter as in the case of the radial distribution functions.  The narrow width of $P(\cos\theta)$ in region~I in Figs.\ \ref{fig4}(c) and~\ref{fig4}(d) confirms that the anions and cations close to the solute (cf.\ Fig.~\ref{fig3}) are situated mainly around the positive and negative sites of the IP solute, respectively.  These ions constitute primarily the first solvation shells corresponding to the first main peaks of the anion and cation radial distributions in Fig.~\ref{fig2}.  In region~II, $P(\cos\theta)$ for both cations and anions are rich in structure with multiple peaks, manifesting that their distributions are highly anisotropic.  It is interesting to note that 
the cation and anion populations in regions I and~II tend to fluctuate in $\cos\theta$ in a staggered fashion, viz., they oscillate out of phase.  This indicates that EMI$^{+}$ rings and Cl$^{-}$ ions tend to alternate in their angular distributions around the IP solute for $r\lesssim 9$\,\AA.  The ionic distributions in region~III are considerably more isotropic than those in regions I and~II.  However, the former are not as isotropic as their counterpart in the presence of NP.  This is another indication that the effects of the solute charge on the solvent structure goes considerably beyond a separation of $\sim 10$\,\AA\ (see below).

The MD results for the solvent structure around the diatomic solutes in EMI$^+$PF$_6^-$ are shown in Fig.~\ref{fig5}.  As expected, EMI$^+$PF$_6^-$ shows essentially the same electrostrictive behavior as EMI$^+$Cl$^-$.  We nonetheless notice that  the first peaks of the anion distributions around the IP and HIP  are located at larger values of $r$ in the former RTIL than in the latter because PF$_6^-$ is bigger than Cl$^-$.  In addition, the anion peak heights  are reduced in EMI$^+$PF$_6^-$ compared with EMI$^+$Cl$^-$.  By contrast, the cation distributions show the opposite trend.  To be specific, $g(r)$ for the M1 site and cation center around the IP and HIP solutes become enhanced in EMI$^+$PF$_6^-$ at both $\rho =1.31$ and 1.375\,g/cm$^{3}$, compared with those in EMI$^+$Cl$^-$ at $\rho =1.1$ and 1.2\,g/cm$^{3}$.  Though not presented here, the distribution of the C2 site (Fig.~\ref{fig:emi}) shows a similar trend.  This implies that in the density range we investigated, the cation contributions to RTIL energetics and dynamics would be more significant in EMI$^+$PF$_6^-$ than in EMI$^+$Cl$^-$. 
We will revisit this point in Sec.~\ref{subsec:free} below and in II.
We also note that the behavior of $N(r)$, Eq.~(\ref{eq:int:density}), in EMI$^+$PF$_6^-$ (not shown here) is similar to that in EMI$^+$Cl$^-$ in Fig.~\ref{fig3ad}.

We now turn to benzenelike solutes.  The cation and anion radial distributions in the presence of RB and DB in EMI$^+$Cl$^-$ and in EMI$^+$PF$_6^-$ are exhibited in Fig.~\ref{fig6}.  Comparison with Figs.\ \ref{fig2} and~\ref{fig5} shows that their basic trends are very similar to the diatomic solute case.  Specifically, both the anion and cation distributions become more structured and their densities very close to the solute become higher as the solute charge separation is increased.  Also as the anion size grows, the cation structure near the solute becomes enhanced, especially in the case of DB.  One noteworthy feature is that in both RTILs, the ratio of the first peak heights of the EMI$^+$ center and anion radial distributions is larger in the presence of the benzenelike DB than that in the presence of the diatomic IP.  For example, in EMI$^+$Cl$^-$, the main anion peak around IP in Fig.~\ref{fig2}(a) is about 4--5 times higher than the corresponding cation peak in Fig.~\ref{fig2}(c), while the former [Fig.~\ref{fig5}(a)] is only twice as large as the latter [Fig.~\ref{fig5}(b)] in the presence of DB.  In the case of EMI$^+$PF$_6^-$,  the first peak of EMI$^+$ is actually higher than that of PF$_6^-$ in the presence of DB, whereas it is the opposite in the presence of IP.   Thus, 
the relative enhancement of cation structure with increasing solute charge separation is more pronounced near the benzenelike solute than near the diatomic solute. 
This implies that cations play a more significant role in solvation in the presence of the former solute than in the presence of the latter (see below).

We examine the ring orientation of EMI$^+$ relative to benzene as a function of the angle $\phi$ between the two unit vectors normal to the benzene and imidazolium rings.  The results for EMI$^+$PF$_6^-$ are displayed in Fig.~\ref{fig:ring}.  While the ring orientation of EMI$^+$ in the ``bulk'' region, i.e., more than 5.5\,\AA\ from the solute center, is almost isotropic, those in the first solvation shell show a rather anisotropic distribution with  peaks around $\cos\phi=\pm 0.9$ for RB and $\pm 0.98$ for DB.  This means that relatively parallel configurations of the EMI$^+$ and benzene rings, i.e., $\pi$-stacking, are favored near the solute, consonant with a previous simulation study.\cite{lynden-bell:benzene}
Though not presented here, our analysis shows that  on the average, the M1 and E1 groups of EMI$^+$ (Fig.~\ref{fig:emi}) close to the solute  face outward when viewed from the center of the solute.  This allows a closer approach of the cations to the solute by avoiding the steric hindrance between the two alkyl groups and benzene ring.

Finally, we investigate the screening behavior of electric charges in EMI$^+$Cl$^-$
and EMI$^+$PF$_6^-$.  In the absence of a rigorous theoretical framework to describe  RTILs, we employ the formulation developed for simple ionic systems. 
In the strongly-coupled regime, the charge distribution around ionic species $\alpha$ at a large separation takes the form\cite{screen,molten_salt:sim,screen2}
\begin{equation}
Q_\alpha(r)= \frac{A_\alpha}{r} \mathrm{e}^{- r/\lambda} \sin\left({\frac{2\pi r}{d}+\varphi}\right)\ ,
\label{eq:screen}
\end{equation} 
where $A_\alpha$ describes the $\alpha$-dependence, $\lambda$ is the screening length, and $d$ and $\varphi$ are the period and phase shift associated with charge oscillations in the $r$-space, respectively.  We examine $Q_\alpha(r)$ with respect to both the anionic species and various charge sites of EMI$^+$.  For illustration, the MD results for $\ln |rQ_\alpha(r)|$ for $\alpha = \mathrm{Cl^{-}}$ and C2 site of the cation (Fig.~\ref{fig:emi}) in EMI$^+$Cl$^-$ at density $\rho=1.1$\,g/cm$^{3}$ are plotted in Fig.~\ref{fig:screen}.  Despite the limitation in the $r$-range probed via MD in the present study, we observe two notable features there.  First, to a good approximation, the $r$-dependence of the envelop of $|rQ_\alpha(r)|$ displays an exponential decay except at short separations.  In fact, for $r\gtrsim 6$\,\AA, Eq.~(\ref{eq:screen}) apparently provides quite an accurate description for the charge distributions in RTILs, including oscillations (see below).  The respective screening lengths for $\alpha = \mathrm{Cl^{-}}$ and C2 are around 9 and 10\,\AA.  A similar analysis for the M1 and N1 sites of EMI$^+$ yields $\lambda\approx 7$ and 9\,\AA, respectively.  We thus conclude that the screening length in EMI$^+$Cl$^-$ at $\rho=1.1$\,g/cm$^{3}$ is in the range\ 7--10\,\AA.  For EMI$^+$PF$_6^-$ at 1.31\,g/cm$^{3}$, we obtain $\lambda=10$--20\,\AA.  Though numerical uncertainties associated with the latter are somewhat large, our results undoubtedly demonstrate that the screening length is of the order of 10\,\AA\ for both RTILs.  We also notice that electric charges are better screened in EMI$^+$Cl$^-$ with small anions than in EMI$^+$PF$_6^-$ with large anions.  For perspective, we make brief contact with the
well-known Debye-H{\"u}ckel theory,\cite{dh,dh:review} which yields
\begin{equation}
Q_\alpha(r)= \frac{A_\alpha}{r} \mathrm{e}^{- r/\lambda_{\text{DH}}} 
\label{eq:dh}
\end{equation}
with the screening length
$\lambda_{\text{DH}}=(4\pi \sum_\alpha n_\alpha q_\alpha^2)^{-1} k_{\text{B}} T$.
Its direct application to the RTILs would yield $\lambda_{\text{DH}}\approx 0.15$~\AA\ and 0.18~\AA\ for EMI$^+$Cl$^-$ and EMI$^+$PF$_6^-$, respectively.  The Debye-H{\"u}ckel theory, which completely ignores molecular size effects and is thus valid only in the infinite dilution limit of ionic solutions, overestimates the ionic screening behavior of RTILs by about two orders of magnitude.\cite{keblinski}

The second aspect we point out is that as alluded to above, the charge distribution oscillates quite regularly in Fig.~\ref{fig:screen}.  The respective $d$ values for EMI$^+$Cl$^-$ and EMI$^+$PF$_6^-$ are about 5.7\,\AA\ and 6.6\,\AA.  This seems to indicate that the distribution of concentric (but smeared-out) shells of alternating charges around a central ion, a picture originally put forward for molten salts,\cite{molten_salt} also applies to the present RTILs.  While this may be the case for spherical anions, we expect that charge distributions around nonspherical EMI$^+$ with an extended molecular structure would be rather aniosotropic.

The radial charge distributions $Q_{\pm}(r)$ around the positive and negative sites of the IP solute in  EMI$^+$Cl$^-$ are exhibited in Fig.~\ref{fig:screen_DP}.  Analogous to the charge distributions around the solvent ions in Fig.~\ref{fig:screen}, $Q_{\pm}(r)$ show oscillations in $r$ with a period $d\approx 6$\,\AA.  This alternation of cation-rich and anion-rich regions around IP was already noted above in connection with $N(r)$ in Fig.~\ref{fig3ad}.  We notice that the locations of the $Q_{+}(r)$ minima coincide with the peak positions of the chloride distribution around the (+) site of IP in Fig.~\ref{fig2}(a).  The locations of the $Q_{-}(r)$ maxima generally agree with the peak positions of the cation distributions in Figs.\ \ref{fig2}(b) and~\ref{fig2}(c).\cite{comment:charge}  
For comparison, the solvent charge distribution in the presence of NP is also displayed in Fig.~\ref{fig:screen_DP}.  In direct contrast with the IP case, the radial charge density around NP is essentially neutral with minor fluctuations.  This difference in $Q(r)$ between the NP and IP emphasizes the dramatic effect the solute charges have on solvent structure and related charge distributions in ionic liquids.  We also notice that the IP charges are screened by the solvent just like the ions.  The screening lengths for $Q_{\pm}(r)$ appear to be similar to those associated with the solvent ions,  viz., $\sim\!10$\,\AA\ (although we would need considerably improved MD statistics to accurately quantify the former).  This provides an explanation for the relatively long range of the solute charge effect, i.e., the structure of the solvent ions located more than 10\,\AA\ from the solute is still significantly affected by the solute charge distributions.  We note that the screening of the solute charges is invoked to explain the non-fluctuating behavior of the average solute-solvent electrostatic interactions at large separations in Ref.~\onlinecite{kobrak}.

\subsection{Solvation Free Energetics}
\label{subsec:free}

We proceed to free energetic aspects of solvation in RTILs, especially equilibrium solvation stabilization and fluctuations relevant to charge transfer and shift processes.  For simplicity, we assume that the solute is characterized by two nonpolarizable electronic states $a$ and $b$, which are degenerate in energy.  For a given solvent configuration ${\cal Q}$ and a solute electronic state $i$ ($=a,b$), we denote the total energy of the combined solute-solvent system as $E_i({\cal Q})$. 
The Franck-Condon (FC) energy $\Delta E_{a\to b}({\cal Q})$ associated with the $a\to b$ transition in the presence of ${\cal Q}$ is given by
\begin{equation}
\Delta E_{a\to b}({\cal Q})= E_a({\cal Q}) - E_b({\cal Q})\ .
\label{eq:egap}
\end{equation}
Since Eq.~(\ref{eq:egap}) measures the solvation stabilization difference between states $a$ and $b$, which varies with the solvent configuration ${\cal Q}$, the FC energy $\Delta E_{a\to b}$ provides a very convenient variable (``solvent coordinate'') that can be used in lieu of ${\cal Q}$ to describe the collective influence of the solvent on the solute.\cite{warshel}  
Since all intermolecular interactions are assumed to be pairwise additive and the solute LJ parameters do not vary with the solute electronic states, only the Coulombic interactions between the solute and solvent contribute to $\Delta E_{a\to b}$ in Eq.~(\ref{eq:egap}) in our description. 

The probability distribution $P_{a/b}$ of the FC energy $\Delta E_{a\to b}$ when the solvent is in equilibrium with the $a$-state solute is given by
\begin{equation}
P_{a/b}(\Delta E_{a\to b}) = \int d{\cal Q}\, f_a^{\text{eq}}({\cal Q}) \,
\delta (\Delta E_{a\to b}-\Delta E_{a\to b}({\cal Q}))\ .
\label{eq:probability}
\end{equation}
Here $f_a^{\text{eq}}({\cal Q})$ is the equilibrium ensemble distribution of ${\cal Q}$ in the presence of the $a$-state solute and $\delta(\cdots)$ is the Dirac delta function. The equilibrium average and fluctuations of $\Delta E_{a\to b}({\cal Q})$ then read
\begin{eqnarray}
\langle\Delta E_{a\to b}\rangle &=& \int dx\, x\, P_{a/b}(x) \nonumber\\
\langle(\delta\Delta E_{a\to b})^2\rangle &=& \int dx\, x^2\, P_{a/b}(x) -
\langle\Delta E_{a\to b}\rangle^2\ ,
\label{eq:average}
\end{eqnarray}
where $\delta\Delta E_{a\to b}$ is the deviation of $\Delta E_{a\to b}$ from its equilibrium average, i.e., $\delta\Delta E_{a\to b} \equiv \Delta E_{a\to b}-\langle\Delta E_{a\to b}\rangle$.
We also introduce a free-energy function for solvent fluctuations:
\begin{equation}
F_{a/b}(\Delta E_{a\to b}) \equiv - k_B T \ln P_{a/b}(\Delta E_{a\to b})\ .
\label{eq:free:energy}
\end{equation}
This defines an effective electronic potential energy curve, upon which solvation processes of the $a$-state solute, monitored via a FC transition to the $b$-state, occur.  
With the aid of the equipartition principle, we can determine the force constant $k_{a/b}$ associated with $F_{a/b}$ according to\cite{carter:hynes}
\begin{equation}
k_{a/b} = \frac{k_B T}{\langle(\delta\Delta E_{a\to b})^2\rangle}\ .
\label{eq:force}
\end{equation}
The underlying assumption in Eq.~(\ref{eq:force}) is that $P_{a/b}(\Delta E_{a\to b})$ defined in Eq.~(\ref{eq:probability}) is gaussian and thus $F_{a/b}(\Delta E_{a\to b})$ is harmonic.

Since $\langle\Delta E_{a\to b}\rangle$ is the (free) energy difference between solute state $a$ in equilibrium with the solvent and its FC-(de)excited state $b$ and the two states are degenerate in energy in vacuum, it describes the energy shift in the ${a \to b}$ transition induced by solvation in RTILs.  Therefore, $\langle\Delta E_{a\to b}\rangle$ with dipolar $a$-state is similar to various empirical polarity scales based on solvent spectral shifts, which gauge the solvation power of RTILs.  To quantify the effective polarity of the RTILs, we have computed $\langle\Delta E_{\text{IP}\to\text{NP}}\rangle$ and  $\langle\Delta E_{\text{DB}\to\text{RB}}\rangle$ for EMI$^+$Cl$^-$ and for EMI$^+$PF$_6^-$ and compiled the results, together with those for acetonitrile for comparison, in Table~\ref{table:result}.  It is found that the effective polarity of these RTILs is very high.  In fact, their values for $\langle\Delta E_{\text{IP}\to\text{NP}}\rangle$ are larger than that of highly dipolar acetonitrile, indicating that as far as solvation stabilization is concerned, EMI$^+$Cl$^-$ and EMI$^+$PF$_6^-$ behave as more ``polar'' solvents than $\mathrm{CH_3CN}$.  This is in good agreement with recent spectroscopic measurements\cite{ref2,pandey}
because 1-butyl-3-methylimidazolium hexafluorophosphate (BMI$^{+}$PF$_6^-$) and EMI$^+$PF$_6^-$ are similar in their effective polarity (see below).  We also observe that the values of $\langle\Delta E_{\text{DB}\to\text{RB}}\rangle$  are smaller than the corresponding $\langle\Delta E_{\text{IP}\to\text{NP}}\rangle$ values by more than 10\,\%.  We believe that the difference in the degree of charge separation between the diatomic and benzenelike solutes is mainly responsible for this.  Since the dipole moment difference of the DB and RB solutes (13.5\,D) is smaller than that of the IP and NP solutes (16.7\,D), the associated solvation stabilization difference is smaller for the former solute type than the latter.  We also notice that the value of $\langle\Delta E_{\text{IP}\to\text{NP}}\rangle$ and thus effective polarity increase with the solvent density. 

Another noteworthy feature about $\langle\Delta E_{a\to b}\rangle$ is that its values for
$\mathrm{IP\to NP}$ and $\mathrm{DB\to RB}$ are lower in EMI$^+$PF$_6^-$ than in EMI$^+$Cl$^-$.  This suggests that for a given cationic species, the solvating power of RTILs decreases as their anion size grows.  
To gain further insight, we have analyzed the cation and anion contributions, denoted by $\langle\Delta E_{a\to b}^{\text{cat}}\rangle$ and $\langle\Delta E_{a\to b}^{\text{an}}\rangle$, to $\langle\Delta E_{a\to b}\rangle$.  The results in Table~\ref{table:result} show that the anions play a major role in solvation stabilization in EMI$^+$Cl$^-$.  For both $a=\mathrm{IP}$ and DB, $\langle\Delta E_{a\to b}^{\text{an}}\rangle$ arising from Cl$^{-}$ is larger than $\langle\Delta E_{a\to b}^{\text{cat}}\rangle$ from EMI$^+$ by more than a factor of two.  In EMI$^+$PF$_6^-$, on the other hand, the anion contributions become lowered significantly by 30--40\,\% for $a=\mathrm{IP}$ and by $\sim\!50$\,\% for DB, compared with EMI$^+$Cl$^-$.  This has the origin in the fact that bulky PF$_6^-$ cannot approach the (+) site of the solute as closely as small Cl$^-$ [cf.\ Figs.\ \ref{fig2}(a) and \ref{fig5}(a) above]. 
By contrast, the cation contributions to $\langle\Delta E_{a\to b}\rangle$ increase with the anion size.  We attribute this to the above-mentioned enhancement in cation structure near the solute, especially around its negative site, when Cl$^-$ is replaced by PF$_6^-$ [see Figs.\ \ref{fig2}(b)(c) and~\ref{fig5}(b)(c) above].   As a consequence, the contributions of cations to solvation stabilization and thus to effective solvent polarity become comparable to (and sometimes even exceed) those of anions in EMI$^+$PF$_6^-$.  Nevertheless, effects of common cations on $\langle\Delta E_{a\to b}\rangle$ changes are of secondary importance compared with those of the anions which we vary.   
By analogy we would expect that modulations of the cation size by, e.g., varying its alkyl chain length would have similar effects on the solvation strength and effective polarity of RTILs\cite{ref2} though they would be considerably smaller than those we found here between EMI$^+$Cl$^-$ and EMI$^+$PF$_6^-$. Thus, for example, the effective polarity of BMI$^{+}$PF$_6^-$ with a longer butyl chain would be somewhat lower than that of EMI$^+$PF$_6^-$ with a shorter ethyl chain.

We notice in Table~\ref{table:result} that for a given RTIL, the relative contribution of cations to $\langle\Delta E_{a\to b}\rangle$ compared to that of anions is larger for $\mathrm{DB\to RB}$ (i.e., benzenelike solute) than for $\mathrm{IP\to NP}$ (i.e., diatomic solute).  In EMI$^+$Cl$^-$, for example, $\langle\Delta E_{a\to b}^{\text{cat}}\rangle$ is $\sim\! 29$\,\% of the total spectral shift for $\mathrm{IP\to NP}$, while it accounts for 32\,\%\ in the case of $\mathrm{DB\to RB}$.  In EMI$^+$PF$_6^-$ at 1.375\,g/cm$^3$, the cation contributions for the former and latter are, respectively, 46\,\%\ and 56\,\%.  We believe that this is closely related to differing solvent structure enhancement with the solute type as discussed near Fig.~\ref{fig5} above. To be specific, for the RTILs studied here,  the relative degree of structure making associated with cations is higher near the benzenelike solute than near the diatomic solute as the solute charge separation grows.  Accordingly we expect that the relative contribution to $\langle\Delta E_{a\to b}\rangle$ from cations is larger in the presence of DB than that in the presence of IP.   This suggests that respective roles played by cations and anions in $\langle\Delta E_{a\to b}\rangle$ and related solvation properties are influenced by the solute type---both geometry and charge distribution---mainly through the modulations of the ion distributions near the solute.  This would mean that the relative contributions of anions and cations to the solvent spectral shift in the presence of, e.g., betaine-30 studied in Ref.~\onlinecite{kobrak}, are reduced and enhanced, respectively, compared to the diatomic solute considered here.  This is because the main positive site of the former is more embedded in the solute and thus less exposed to the solvent ions than its negative site.

Turning to solvent fluctuations relevant to solvation in RTILs, we have found that
$\langle(\delta\Delta E_{a\to b})^2\rangle$ diminishes as the solute charge separation is increased (Table~\ref{table:result}).   
This means that the effective potential energy curve $F_{a/b}(\Delta E_{a\to b})$ in Eq.~(\ref{eq:free:energy}) becomes tighter and its force constant $k_{a/b}$ in Eq.~(\ref{eq:force}) larger with the growing solute charge separation (see Fig.~\ref{fig:free}).     
This trend is more pronounced in the presence of benzenelike solutes than diatomic solutes.  We ascribe this to the enhancement of rigidity in solvent structure close to the solute.  As mentioned several times above, electrostriction yields solvent density enhancement near the solute as the charge separation of the latter increases.  We believe that this reduces the free volume available for individual solvent ions and thus suppresses solvent structural fluctuations near the solute.  This in turn lowers $\langle(\delta\Delta E_{a\to b})^2\rangle$, which is directly related to solvent configuration fluctuations. 
It is interesting to note that according to the present and previous studies,\cite{carter:hynes,maroncelli} aprotic solvents such as $\mathrm{CH_3CN}$ and $\mathrm{CH_3Cl}$ show a similar trend in the solvent force constant. 

We add a cautionary remark: The solvent density increase in the proximity of the solute also strengthens the solute-solvent Coulombic interactions, i.e., the magnitude of $\Delta E_{a\to b}$ increases.  Assuming all other aspects of solvation would remain unchanged, we can infer that this enhancement in the electrostatic interaction would {\it increase} $\langle(\delta\Delta E_{a\to b})^2\rangle$.  This is the opposite of the structural rigidity effects mentioned above.  
Thus qualitatively speaking, electrostriction causes two opposing factors to increase with the solute charge separation---viz., solvent structural rigidity and electrostatic interaction strength, which play antagonistic roles in the fluctuations of $\Delta E_{a\to b}$.  For RTILs (and acetonitrile) studied here, the modulation of structural rigidity appears to dominate over that of electrostatic interaction strength, so that  $\langle(\delta\Delta E_{a\to b})^2\rangle$ decreases as the solute charge separation grows.  We parenthetically note that the opposite trend obtains in water, i.e., $\langle(\delta\Delta E_{a\to b})^2\rangle$ increases with the solute charge separation.\cite{comment:force:trend,comment:force:trend2} 
%
%

We next consider the solvent reorganization free energy $\lambda_{a/b}$, which describes the free-energy cost associated with rearranging the solvent on the $F_{a/b}$ curve from the configurations in equilibrium with solute state $a$ to those with solute state $b$.  Since the solvent force constant $k_{a/b}$ varies with the solute electronic state, so does $\lambda_{a/b}$.  If we assume that $F_{a/b}$ is harmonic, the state-dependent reorganization free energy $\lambda_{a/b}$ is given by\cite{comment:nonlinear}
\begin{equation}
\lambda_{a/b} = \frac{1}{2} k_{a/b}\left(\Delta E_{a\to b}^{\text{min}}+\Delta E_{b\to a}^{\text{min}}\right)^2
\approx \frac{1}{2} k_{a/b}\left(\langle\Delta E_{a\to b}\rangle + \langle\Delta E_{b\to a}\rangle\right)^2\ ,
\label{eq:reorg}
\end{equation}
where $\Delta E_{a\to b}^{\text{min}}$ is the value of $\Delta E_{a\to b}$ at the minimum of $F_{a/b}$.\cite{comment:accuracy} 
Using the local force constant determined via Eq.~(\ref{eq:force}), we calculate $\lambda_{a/b}$ in Eq.~(\ref{eq:reorg}) and present the results in Table~\ref{table:result}.  For EMI$^+$Cl$^-$ at $\rho=1.1$\,g/cm$^{3}$, we obtain $\lambda_{\text{IP}/\text{NP}}=66.8$\,kcal/mol and $\lambda_{\text{NP}/\text{IP}}=48.5$\,kcal/mol.  The corresponding values for EMI$^+$PF$_6^-$ at 1.375\,g/cm$^{3}$ are 68.4 and 45.7\,kcal/mol.  This shows that charge shift processes in RTILs such as $\mathrm{IP\leftrightharpoons NP}$ are accompanied by huge solvent reorganization.  Here it is of interest to make comparison with acetonitrile, in which the reorganization free energy is $\lambda_{\text{IP}/\text{NP}}=41.5$ and $\lambda_{\text{NP}/\text{IP}}=36.9$\,kcal/mol at room temperature.  According to recent spectroscopic measurements,\cite{bright,pandey} the reorganization free energy in BMI$^{+}$PF$_6^-$ is comparable to that in acetonitrile.  Since  BMI$^{+}$PF$_6^-$ is expected to be somewhat lower in effective polarity than EMI$^+$PF$_6^-$ (see above), the solvent reorganization energy in the former would be smaller than that in the latter [see Eq.~(\ref{eq:stokes}) below].  This is consistent with our finding that $\lambda_{a/b}$ in EMI$^+$Cl$^-$ and EMI$^+$PF$_6^-$ are in general larger than that in CH$_3$CN.  We also mention that because the free energy curve is tighter in the presence of IP than in the presence of NP, $\lambda_{\text{IP}/\text{NP}}$ is larger than $\lambda_{\text{NP}/\text{IP}}$ and similarly for the DB and RB states.\cite{water}  

In view of the state-dependent solvation behavior observed above, we briefly examine the assumption of harmonic potentials invoked in Eq.~(\ref{eq:reorg}).  The reorganization free energies and FC transition energies satisfy\cite{comment:stokes}
\begin{equation}
\lambda_{a/b} + \lambda_{b/a}
=\Delta E_{a\to b}^{\text{min}} + \Delta E_{b\to a}^{\text{min}}
\approx\langle\Delta E_{a\to b}\rangle + \langle\Delta E_{b\to a}\rangle ,
\label{eq:stokes}
\end{equation}
whether or not $F_{a/b}$ and $F_{b/a}$ are harmonic.  Comparing the predictions for $\lambda_{a/b} + \lambda_{b/a}$ in the harmonic assumption with the MD results for $\langle\Delta E_{a\to b}\rangle + \langle\Delta E_{b\to a}\rangle$,\cite{comment:accuracy} we find easily that Eq.~(\ref{eq:stokes}) holds only approximately.  For instance, the estimate for $\langle\Delta E_{\text{IP}\to \text{NP}}\rangle + \langle\Delta E_{\text{NP}\to \text{IP}}\rangle$ in the harmonic approximation is 115\,kcal/mol in EMI$^+$Cl$^-$ at $\rho=1.1$\,g/cm$^{3}$, while the MD yields 101\,kcal/mol.  The corresponding results for EMI$^+$PF$_6^-$ at 1.375\,g/cm$^{3}$ are 114 and 95\,kcal/mol, respectively.  We thus conclude that the anharmonicity in $F_{a/b}$ plays a non-negligible role in solvation free energetics.  In view of the drastic solvent structure change and related solvent force constant variation with the solute charge distribution, the presence of nonlinearity in solvent response is not surprising.  What is surprising is that despite huge electrostrictive effects, the harmonic approximation for $F_{a/b}$ with the state-dependent force constant\cite{comment:nonlinear} provides a reasonable description to quantitate the solvent reorganization free energy in RTILs.  While a similar harmonic approximation works well for dipolar solvents,\cite{carter:hynes,harmonic_dp} the RTILs are a totally different matter in that they are comprised of free ions.  Related issues for solvent dynamics, e.g., linear response, are considered in II.

Before we conclude, we consider the united-atom description employed in the present study for perspective. 
Because the $\mathrm{CH_3}$ and $\mathrm{CH_2}$ moieties of EMI$^+$ (Fig.~\ref{fig:emi}) are spherically symmetric in their shape and charge distribution in the united-atom description, the directionality
in their interactions with other molecules (i.e., dependence on relative orientations of their H atoms) present in real ionic liquids is completely neglected in our MD.  Also two rigid alkyl groups of EMI$^+$ could introduce spurious hindrance compared with a more accurate flexible description.
As for PF$_6^-$, while the rigid, united-atom representaion may be reasonable for short-range interactions because of its high symmetry, the extended character of its charge distribution ignored in the present study could play a nonnegligible role in, e.g., $\Delta E_{a\to b}$.
To gain insight into thses issues, we have performed a test simulation using the flexible, all-atom description of Ref.~\onlinecite{ref:margulis} for EMI$^+$ and the rigid, all-atom description of Ref.~\onlinecite{kaminski} for PF$_6^-$.  Its results, averaged over a 2~ns trajectory in the presence of the diatomic solute in EMI$^+$PF$_6^-$ are given in Table~\ref{table:result}.  We notice that
the united-atom and all-atom descriptions compare well in their predictions for solvation properties.  In fact, considering the difference in their parametrization,  the degree of agreement between the two in the solvent spectral shift and its anion and cation components, solvent fluctuations and reorganization free energy is quite remarkable.  Though very preliminary, we have found a similar agreement between the united-atom and all-atom descriptions for EMI$^+$Cl$^-$.  This seems to indicate that despite its approximate nature, the united-atom description employed in the present study captures the essential features of RTILs very well.  For related issues on dynamics, the reader is referred to II.

\section{Concluding Remarks}

We have studied equilibrium solvation structure and free energetics in ionic liquids consisting of 1-ethyl-3-methylimidazolium cations paired with either chloride or hexafluorophosphate anions.  We have found that solvent structure varies dramatically with the solute charge distribution.  To be specific, as the solute charge separation increases, both the cation and anion radial distribution functions become more structured and their densities close to the solute become enhanced.  Also their angular distributions around the solute become highly anisotropic except in the bulk region. For instance, the cation and anion populations near the IP solute, though smeared-out, tend to alternate in angular distributions.  
The charge distributions around the solvent ions and IP solute show screened oscillations.  The screening length is estimated to be of the order of 10~\AA\ in both  EMI$^+$Cl$^-$ and EMI$^+$PF$_6^-$ although the statistical error associated with the screening of the IP solute is large.  It was observed that the screening length is not sensitive to the presence of a solute or its charge character.  

We examined the effective polarity of EMI$^+$Cl$^-$ and EMI$^+$PF$_6^-$, measured as solvent spectral shifts that gauge the solvation stabilization of dipolar solutes.  The MD results show that the effective polarity of EMI$^+$Cl$^-$ and EMI$^+$PF$_6^-$ is very high;  
insofar as their solvating strengths are concerned, both RTILs behave as a more polar solvent than acetonitrile, in accord with experiments.\cite{ref2,bright,pandey} 
EMI$^+$Cl$^-$ is more polar of the two, indicating that solvation power decreases with the increasing ion size.  
Related solvent reorganization free energies were found to be in general higher in these RTILs than in acetonitrile.  

We have analyzed respective roles played by the anions and cations in solvation.  While both of them make important contributions to solvent polarity and reorganization free energy, their relative importance tends to vary with the solute type.  To be specific, the relative contribution of anions becomes reduced in the presence of the benzenelike solute, compared with the diatomic solute case.  In parallel with this, the relative enhancement in cation structure near the solute tends to be more pronounced in the presence of the former than in the presence of the latter solute.  This suggests that variations in solute structure and charge distributions have a considerable influence on details of solvation by modulating the ion distributions close to the solute.

To gain insight into the united-atom description employed in the present study, we 
conducted a test simulation with an all-atom description for EMI$^+$PF$_6^-$ and compared with the united-atom results. We obtained good agreement between the two in the solvent spectral shift, its anion and cation contributions, equilibrium solvent fluctuations and reorganization free energy (related issues on solvent dynamics are examined in II).  Very preliminary results for EMI$^+$Cl$^-$ in the all-atom representation show a similar agreement. This indicates that the united-atom description employed here provides a decent framework to simulate RTILs. 

The free-energy curves relevant to solvent fluctuations and associated force constants were found to vary with the solute charge distribution.  To be specific, the free-energy curves become tighter as the solute charge separation grows.  This was attributed to the rigidity enhancement in solvent structure induced by electrostriction.
This demonstrates that a linear solvation scheme is in general not valid for RTILs.  It was found that the harmonic approximation with state-dependent force constants works reasonably well to quantify, e.g., the solvent reorganization free energy.\cite{comment:nonlinear}

Considering the huge electrostrictive effects on the RTIL structure observed in this work, we expect that dynamic properties of solute molecules, e.g., rotational and vibrational energy relaxation times, will vary significantly with the solute charge distribution.  For example, the rotational friction of the solute will grow considerably with its dipole moment analogous to dielectric friction in dipolar solvents.  A similar trend is also expected  for the solute vibrational energy relaxation.  An analysis of these properties will be reported elsewhere.\cite{rtil:rot:vib}

In the following paper,\cite{rtil:shim2} we consider equilibrium and nonequilibrium solvation dynamics in RTILs.

\acknowledgments
This work was supported in part by the Ministry of Education of Korea through the BK21 Program,
by KOSEF Grant No. 01-2002-000-00285-0, and by NSF Grant No.\ CHE-0098062. 

\begin{thebibliography}{<100>}
\bibitem{comment:phosphonium}
 Phosphonium-based RTILs generally do not show this behavior.  See Ref.~\onlinecite{maroncelli:phosphonium}.
\bibitem{comment:solute:parameter} 
These parametric values for the diatomic solute are  the same as in Ref.~\onlinecite{shim}.
\bibitem{comment:dh}
The Debye-H{\"u}ckel theory, which is valid only in the infinite dilution limit of ionic solutions, predicts monotonically decreasing $Q_\alpha(r)$
\begin{equation}
Q_\alpha(r)= \frac{A_\alpha}{r} \mathrm{e}^{- r/\lambda_{\text{DH}}} \ ;\quad
\lambda_{\text{DH}}=\frac{k_{\text{B}} T}{4\pi \sum_\alpha n_\alpha q_\alpha^2}\ .
\end{equation} 
Its direct application to the RTILs would yield
$\lambda_{\text{DH}}\approx ??$~\AA.
\bibitem{comment:angle}
For example, $\theta = 0^\circ$ and 180$^\circ$ correspond to the directions of the positive and negative sites of the solute from its center, respectively.
\bibitem{comment:charge}
From these observations and from Figs.\ \ref{fig4}(c) and~(d), we can easily imagine that the angular charge distributions around IP, though not studied here, are quite anisotropic.
\bibitem{comment:force:trend}
This was first noted in Ref.~\onlinecite{carter:hynes}.
\bibitem{comment:force:trend2}
 The opposite $k_{a/b}$ trend in water  can be understood as follows.
The effective size of hydrogen in water molecules is small; i.e., the molecular shape of water based on its LJ parameters is quite spherical.  Thus as the local solvent density increases due to electrostriction, fluctuations in orientations of water molecules near the solute via librational motions are not as sterically hindered as non-spherical molecules such as acetonitrile.  Furthermore, the electrostatic interaction strength variation in water would be larger due to (partial) hydrogen bonding formation with dipolar solutes than that in aprotic solvents.  Therefore the electrostatic interaction strength is a dominant factor for  the $\langle(\delta\Delta E_{a\to b})^2\rangle$ trend in water.
\bibitem{comment:noneq:relax}
Thus in $S_{a/b}(t)$, the time-evolution of ${\cal Q}_t$ occurs in the presence of the solute $a$-state, while the initial ${\cal Q}_0$ is equilibrated to state $b$.
\bibitem{comment:accuracy}
Within the numerical accuracy of 1~kcal/mol which is the bin size employed for $\Delta E_{a\to b}$ to calculate $P_{a/b}$ and $F_{a/b}$ (Eqs.\ (\ref{eq:probability}) and~(\ref{eq:free:energy})), the MD results for $\Delta E_{a\to b}^{\text{min}}$ and $\langle\Delta E_{a\to b}\rangle$ are the same.
\bibitem{comment:nonlinear}
The nonlinear solvation effects are captured to some extent through the state-dependent local force constant.
\bibitem{water}
In view of the discussions above, it would be the opposite in the case of water, i.e.,
$\lambda_{\text{IP}/\text{NP}}$ is smaller than $\lambda_{\text{NP}/\text{IP}}$.
\bibitem{comment:solvation:shell}
The first solvation shells for anions and cations (and their corresponding bulk regions) are defined with respect to the positive and negative site of the dipolar solute, respectively.
\bibitem{comment:firstshell}
These ions mainly correspond to the anions and cations of the first solvation shells, which are defined with respect to the positive and negative sites of the solute.
\bibitem{comment:cation:rot}
By decomposing the cation motions into rotational and translational components, we have investigated their respective contributions to $\Delta E$ relaxation.  We have found that the contribution from rotational motions is considerably larger than that from translations.
\bibitem{comment:flex}
However, there is a caveat:  the internal motions of cations are totally absent in the rigid model description we employed.  In Sec.~\ref{subsec:eq} above, we observed that the initial relaxation of $C_{a/b}(t)$ [Fig.~\ref{fig:allatom}] of the flexible model during the first $\sim\!50~fs$ is extremely fast compared to that of the rigid model and attributed it 
to these internal motions, in particular, bending and torsional motions of charge and polar groups.  By analogy, we believe that $S_{a/b}^{\text{cat}}(t)$ and thus $S_{a/b}(t)$ should have a similar ultrafast component, which is not captured in the current rigid description.
\bibitem{keblinski}
Similar results were also obtained in the MD simulations of NaCl.
See Ref.~\onlinecite{molten_salt:sim}.
\bibitem{comment:local:charge}
(a) The total charge inside a similar sphere centered at the negative site of IP is $+1.9e$. (b) The corresponding charge inside a similar region around the positive site is $-1.7e$.
\end{thebibliography}

\newpage
\begin{table}
\caption{Lennard-Jones parameters and partial charges}
\vspace*{10pt}
\begin{tabular}{|c|c|c|c|c|}
\hline
&\ atom\ &\ $\sigma_{ii}$\,($\textrm{\AA}$)\ &\ $\epsilon_{ii}$\,(kJ/mol)\ &\ $q_i$\,($e$)\\
\hline
EMI$^+$$^{\,a)}$\ &\ N\ &\ 3.250\ &\  0.7113\ &\ -0.2670\\
&\ C2\ &\ 3.400\ &\  0.3598\ &\  0.4070\\
&\ N\ &\ 3.250\ &\  0.7113\ &\ -0.2670\\
&\ C4\ &\ 3.400\ &\  0.3598\ &\  0.1050\\
&\ C5\ &\ 3.400\ &\  0.3598\ &\  0.1050\\
&\ M1\ &\ 3.905\ &\  0.7330\ &\  0.3160\\
&\ E1\ &\ 3.800\ &\  0.4943\ &\  0.2400\\
&\ M3\ &\ 3.800\ &\  0.7540\ &\  0.0760\\
&\ H2\ &\ 2.421\ &\  0.0628\ &\  0.0970\\
&\ H4\ &\ 2.421\ &\  0.0628\ &\  0.0940\\
&\ H5\ &\ 2.421\ &\  0.0628\ &\  0.0940\\
\hline
Cl$^{-}$\ &\ &\ 4.400\ &\  0.4190\ &\  -1.000\\
PF$_6^{-}$\ &\ &\ 5.600\ &\  1.6680\ &\  -1.000\\
\hline
diatomic\ &\ NP\ &\ 4.000\ &\  0.8298\ &\ 0.0000\\
 &\ IP\ &\ 4.000\ &\  0.8298\ &\ $\pm$1.0000\\
\hline
benzene$^{\,b)}$\ &\ C\ &\ 3.500\ &\  0.3352\ &\ -0.1350\\
 &\ H\ &\ 2.500\ &\  0.2094\ &\ 0.1350\\
\hline
\end{tabular}
\label{table1}
\end{table}
\vspace*{-20pt}

$^{a)}$~The LJ parameters are from Ref.~\onlinecite{ref6} and partial charges are from Ref.~\onlinecite{ref7}. 



$^{b)}$~Ref.~\onlinecite{benzene}.

\newpage

\begin{table}
\caption{MD results$^{\,a)}$}
\vspace*{10pt}
\begin{tabular}{|cc|c|c|c|c|c|c|}
\hline
solvent\ &\  density\ &\ $a/b$\ &\ $\langle \Delta E_{a\rightarrow b}\rangle$\ &\ $\langle \Delta E_{a\rightarrow b}^{\text{an}}\rangle$\ &\ $\langle \Delta E_{a\rightarrow b}^{\text{cat}}\rangle$\ &\ $\langle (\delta \Delta E_{a\to b})^{2} \rangle$\ &\ $\lambda_{a/b}^{\ b)}$\\
\hline
EMI$^+$Cl$^-$\ &\ 1.1\ &\ NP/IP\ &\ 0.25\ &\ $-0.04$\ &\ 0.29\ &\ 83.3\ &\ 48.5\\
  &\ &\ IP/NP\ &\ 100.6\ &\ 71.1\ &\ 29.5\ &\ 60.5\ &\ 66.8\\
EMI$^+$Cl$^-$\ &\ 1.1\ &\ RB/DB\ &\ $-0.41$\ &\ $-0.69$\ &\ 0.28\ &\ 77.5\ &\ 38.4\\
  &\ &\ DB/RB\ &\ 86.9\ &\ 59.3\ &\ 27.6\ &\ 54.2\ &\ 54.9\\
EMI$^+$Cl$^-$\ &\ 1.2\ &\ NP/IP\ &\ $-0.5$\ &\ $-1.65$\ &\ 1.15\ &\ 72.3\ &\  56.1\\
  &\ &\ IP/NP\ &\ 101.5\ &\ 71.7\ &\ 29.8\ &\ 57.9\ &\ 70.0\\
EMI$^+$PF$_6^-$\ &\ 1.31\ &\ NP/IP\ &\ 0.20\ &\ 0.43\ &\ $-0.23$\ &\ 79.8\ &\ 35.1\\
 &\ &\ IP/NP\ &\ 83.8\ &\ 43.7\ &\ 40.1\ &\ 64.5\ &\ 43.5\\
\ EMI$^+$PF$_6^-$$^{\,c)}$\ &\ 1.34\ &\ NP/IP\ &\ $-0.44$\ &\ $-1.02$\ &\ 0.58\ &\ 80.9\ &\ 37.3\\
 &\ &\ IP/NP\ &\ 87.6\ &\ 48.3\ &\ 39.3\ &\ 61.2\ &\ 49.3\\
EMI$^+$PF$_6^-$\ &\ 1.375\ &\ NP/IP\ &\ 0.53\ &\ 0.60\ &\ $-0.07$\ &\ 77.8\ &\ 45.7\\
 &\ &\ IP/NP\ &\ 94.1\ &\ 51.0\ &\ 43.1\ &\ 52.0\ &\ 68.4\\
EMI$^+$PF$_6^-$\ &\ 1.375\ &\ RB/DB\ &\ 0.15\ &\ 0.28\ &\ $-0.13$\ &\ 81.2\ &\ 25.5\\
 &\ &\ DB/RB\ &\ 72.0\ &\ 31.8\ &\ 40.2\ &\ 38.7\ &\ 53.5\\
CH$_3$CN$^{\,d)}$\ &\ 0.73\ &\ NP/IP\ &\ 0.46\ &\ &\ &\ 46.4\ &\ 38.0\\
 &\ &\ IP/NP\ &\ 76.5\ &\ &\ &\ 41.3\ &\ 42.7\\
\hline
\end{tabular}
\label{table:result}
\end{table}

\vspace*{-20pt}
$^{a)}$~Units for energy: kcal/mol.  Due to the difference in MD statistics, some results here are somewhat different from those in Ref~\onlinecite{shim}. 

$^{b)}$ Estimation in the harmonic approximation with Eq.~(\ref{eq:reorg}).

$^{c)}$ Preliminary results obtained with an all-atom description.  See the text for details.

$^{d)}$ MD results at $T=300$\,K.  The LJ parameters and partial charges employed are from Ref.~\onlinecite{acetonitrile}.

\newpage

\begin{figure}
\caption{Structure of the 1-ethyl-3-methylimidazolium cation.  In simulations, the united atom representation is employed for M1, E1 and M3 moieties.}
\label{fig:emi}
\caption{Radial distributions of cations and anions around the diatomic solute in EMI$^+$Cl$^-$ at $\rho=1.1$\,g/cm$^{3}$: (a) Cl$^-$ around the solute $(+)$ site; (b) M1 around the solute $(-)$ site; (c) EMI$^+$ center around the solute $(-)$ site.  Here $r$ is the solute-ion distance in units of \AA~ and the midpoint of two N sites in the imidazolium ring (cf.\ Fig.~\ref{fig:emi}) is employed as the EMI$^+$ center in (c). The results for $g(r)$ at higher density $\rho=1.2$\,g/cm$^{3}$ are almost the same as those at $\rho=1.1$\,g/cm$^{3}$ shown here.}
\label{fig2}
\caption{Number of (a) Cl$^-$ and (b) EMI$^+$ ions surrounding the diatomic solute in EMI$^+$Cl$^-$ at $\rho=1.1$\,g/cm$^{3}$. The dotted and dashed lines represent, respectively, $N(r)$ around the positive and negative sites of IP, while the solid line denotes that around the neutral atoms of NP.  In (b), the EMI$^+$ center is used in the calculations of $N(r)$.}
\label{fig3ad}
\caption{Radial distributions of cations and anions with respect to the center of the diatomic solute in EMI$^+$Cl$^-$ at $\rho=1.1$\,g/cm$^{3}$: (a) Cl$^-$ and (b) EMI$^+$ center. }
\label{fig3}
\caption{Angular probability distributions of (a) anions and (b) cations around the NP solute and (c) anions and (d) cations around the IP solute in EMI$^+$Cl$^-$ at $\rho=1.1$\,g/cm$^{3}$.  Solid, dashed and dotted lines denote the probability distributions in regions I, II, and III, respectively, while $\theta$ is the angle between the IP dipole vector and the solute-to-ion center-to-center direction.  In (a) and (b), $P(\cos\theta)$ for region III is nearly the same as that for region II and thus is not shown.}
\label{fig4}
\caption{Radial distributions of (a) anions around the $(+)$ site, (b) M1 around the $(-)$ site and (c) EMI$^+$ center around the $(-)$ site of the diatomic solute in EMI$^+$PF$_6^-$ at $\rho=1.375$\,g/cm$^{3}$.  Radial distributions at lower density $\rho=1.31$\,g/cm$^{3}$ are somewhat less structured but their essential features including the peak positions remain nearly the same.}
\label{fig5}
\end{figure}
\begin{figure}
\caption{Radial distributions of solvent ions around the carbon sites of the benzenelike solute.
The results for anions and the center of cations in EMI$^+$Cl$^-$ are presented in (a) and (b), respectively, and the corresponding distributions in EMI$^+$PF$_6^-$ are shown in (c) and (d).   In the case of DB, the anion and cation distributions around only those C atoms that are, respectively, more positive and negative in their charge assignments than the remaining four are shown here.  The solvent densities are the same as in Figs.\ \ref{fig2} and~\ref{fig5}.}
\label{fig6}
\caption{Probability distributions of the EMI$^+$ ring orientation with respect to the benzenelike solute in EMI$^+$PF$_6^-$ at $T=400$\,K and $\rho=1.375$\,g/cm$^{3}$.  $\phi$ is the angle between the normal vectors of the imidazolium and benzene rings.  
The solid and dashed lines represent $P(\cos\phi)$ in the first solvation shells of DB and RB, respectively, which consist of the cations with the center-to-center distance of less than 5.5~\AA\  from the solute.  The rest are defined as the ``bulk'' region, $P(\cos\phi)$ of which in the presence of RB is plotted as a dotted line. $P(\cos\phi)$ is normalized to unity in each of the two regions.
The distribution of the DB ring orientation in the bulk is nearly the same as that of RB and thus is not shown here.}
\label{fig:ring}
\caption{Absolute value of $rQ_{\alpha} (r)$~($\alpha=$Cl$^-$, C2) in EMI$^+$Cl$^-$ at $\rho=1.1$\,g/cm$^3$ as a function of the radial distance $r$ in the presence of NP.  The dotted lines have the slopes $-0.115$ and $-0.10$ for C2 and Cl$^-$, respectively, corresponding to the inverse screening length $\lambda^{-1}$.  Though not presented here, comparison with other cases---specifically, pure RTILs and RTIL solutions containing IP---shows that the screening length and oscillation period are rather insensitive to the presence of  the diatomic solute and its charge character.}
\label{fig:screen}
\caption{
Radial charge distributions $Q_{\pm}(r)$ around the positive (---) and negative ($--$) site of the IP solute in EMI$^+$Cl$^-$ at $\rho=1.1$\,g/cm$^3$. }
\label{fig:screen_DP}
\caption{Free energy curves in the presence of (a) a diatomic and (b) a benzenelike solute in EMI$^+$Cl$^-$ at $\rho=1.1$\,g/cm$^{3}$.  Both the free energy $F_{a/b}$ and deviation $\delta\Delta E_{a\to b}$ of $\Delta E_{a\to b}$ from its average are given in units of kcal/mol.}
\label{fig:free}
\end{figure}

\newpage
\newpage
\vspace*{2in}
\centering
\includegraphics[width=5in]{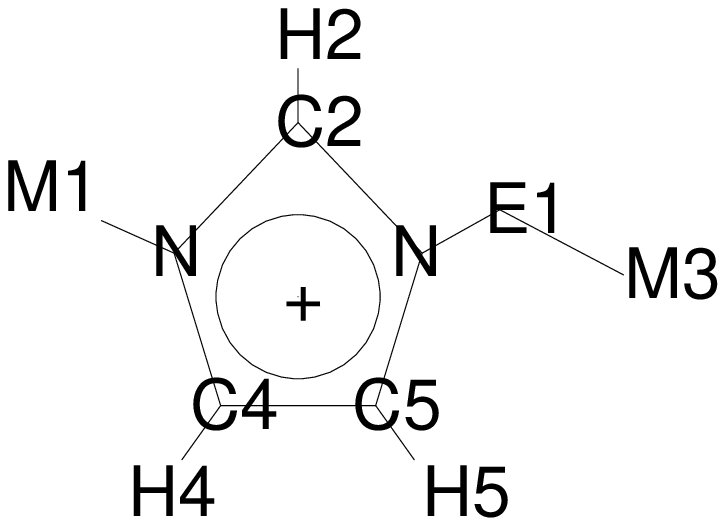}
\centerline{Fig.~\ref{fig:emi}}

\newpage

\centering 
\begin{minipage}{8.0cm} 
\centering 
\includegraphics[width=3.6in]{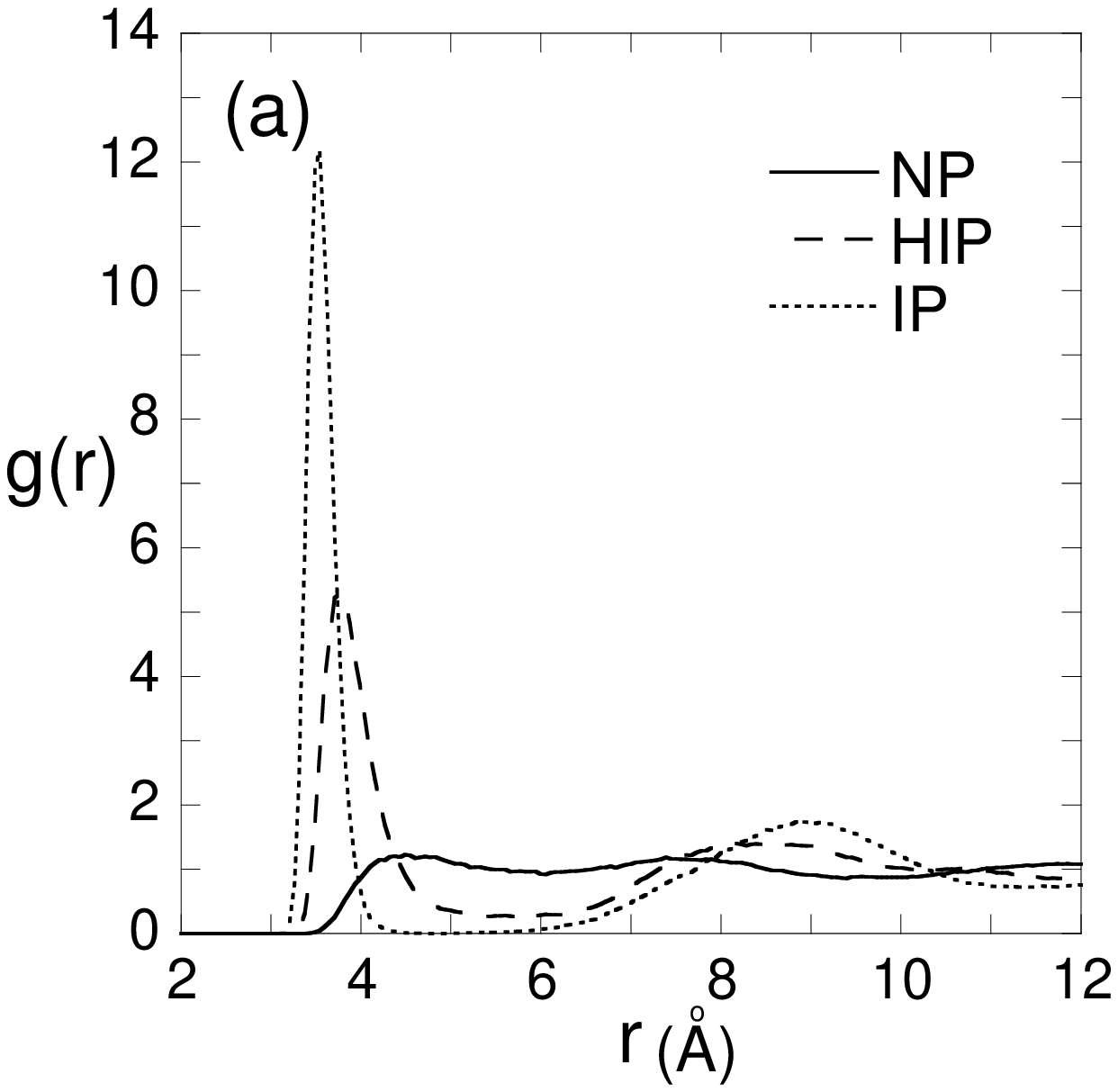} 
\end{minipage} 
\begin{minipage}{8.0cm} 
\centering 
\includegraphics[width=4in]{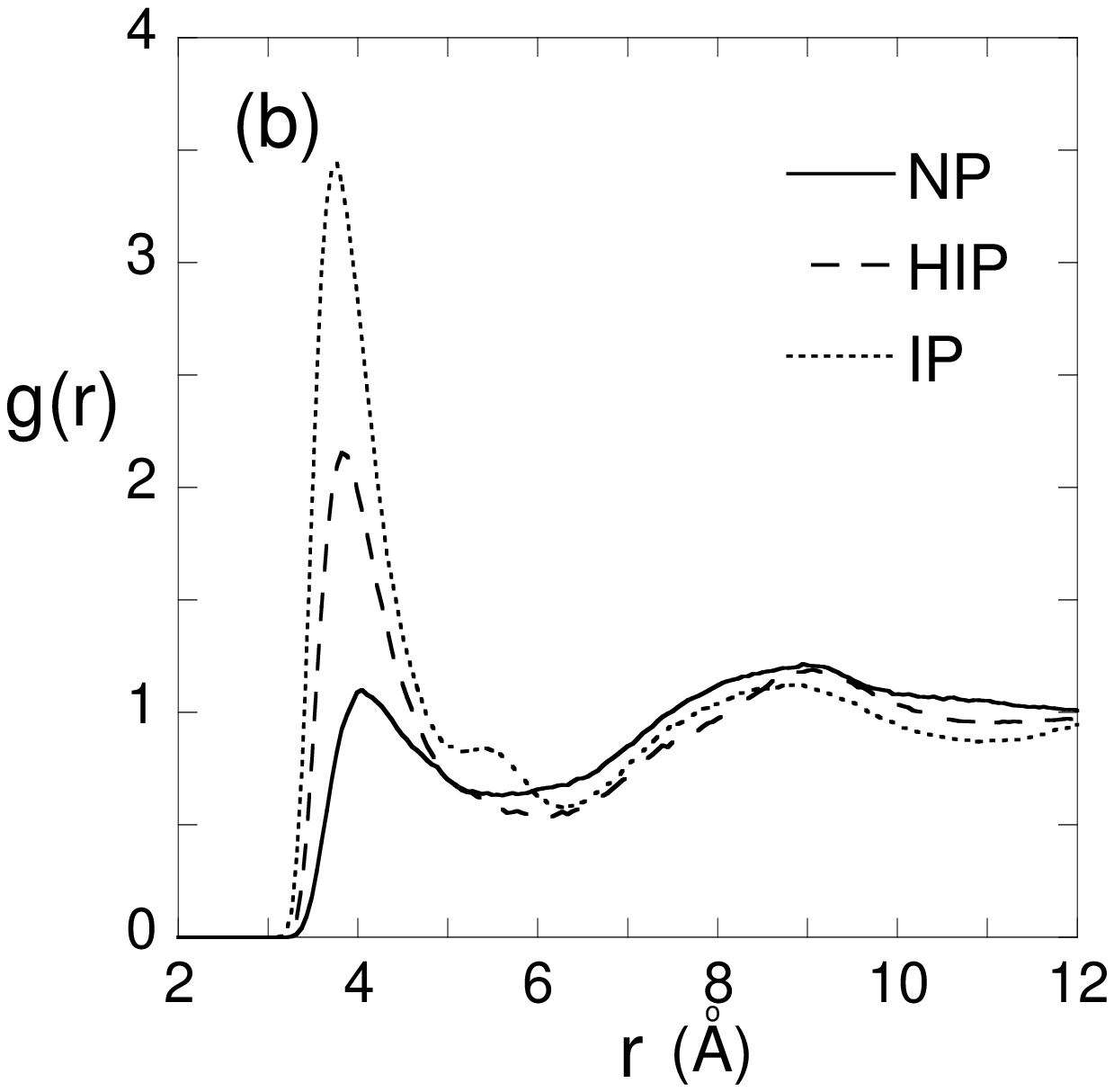} 
\end{minipage} \\[0.5cm] 
\begin{minipage}{8.0cm} 
\centering 
\includegraphics[width=4in]{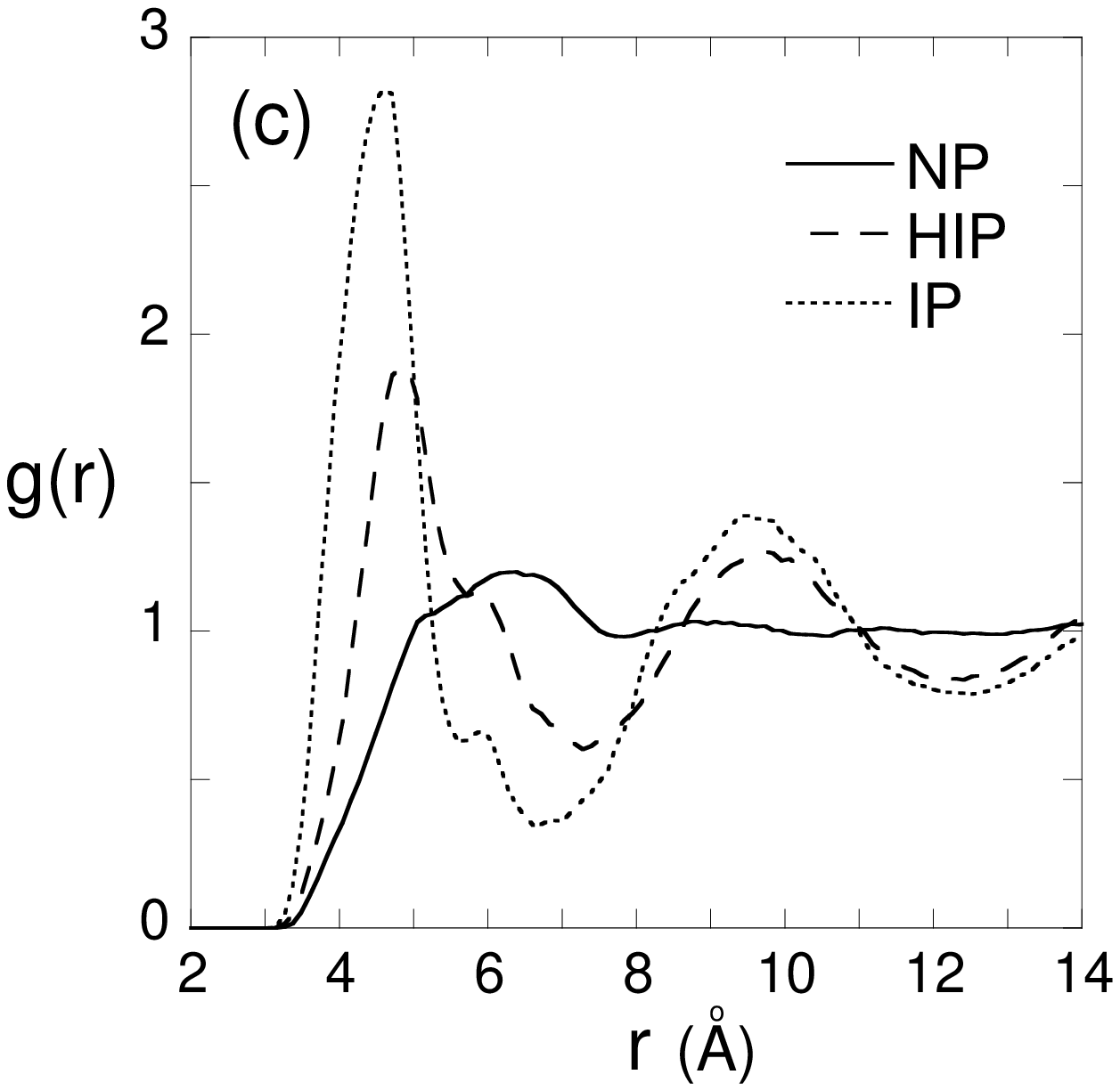} 
\end{minipage} 
\vspace{10cm}
\centerline{\centerline{Fig.~\ref{fig2}}}

\newpage

\centering
\includegraphics[width=4.0in]{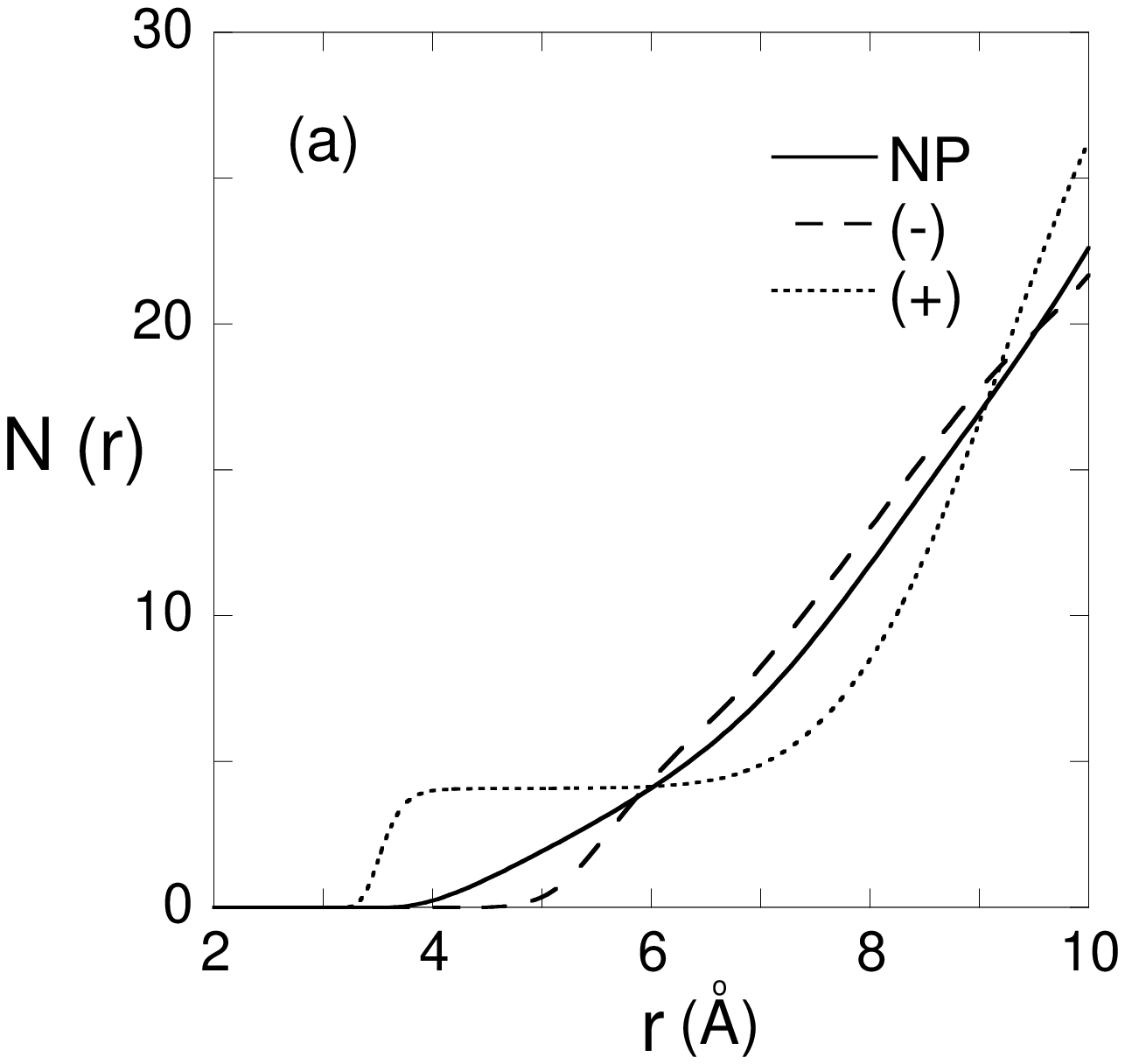}\vspace{0.5in}
\includegraphics[width=4.0in]{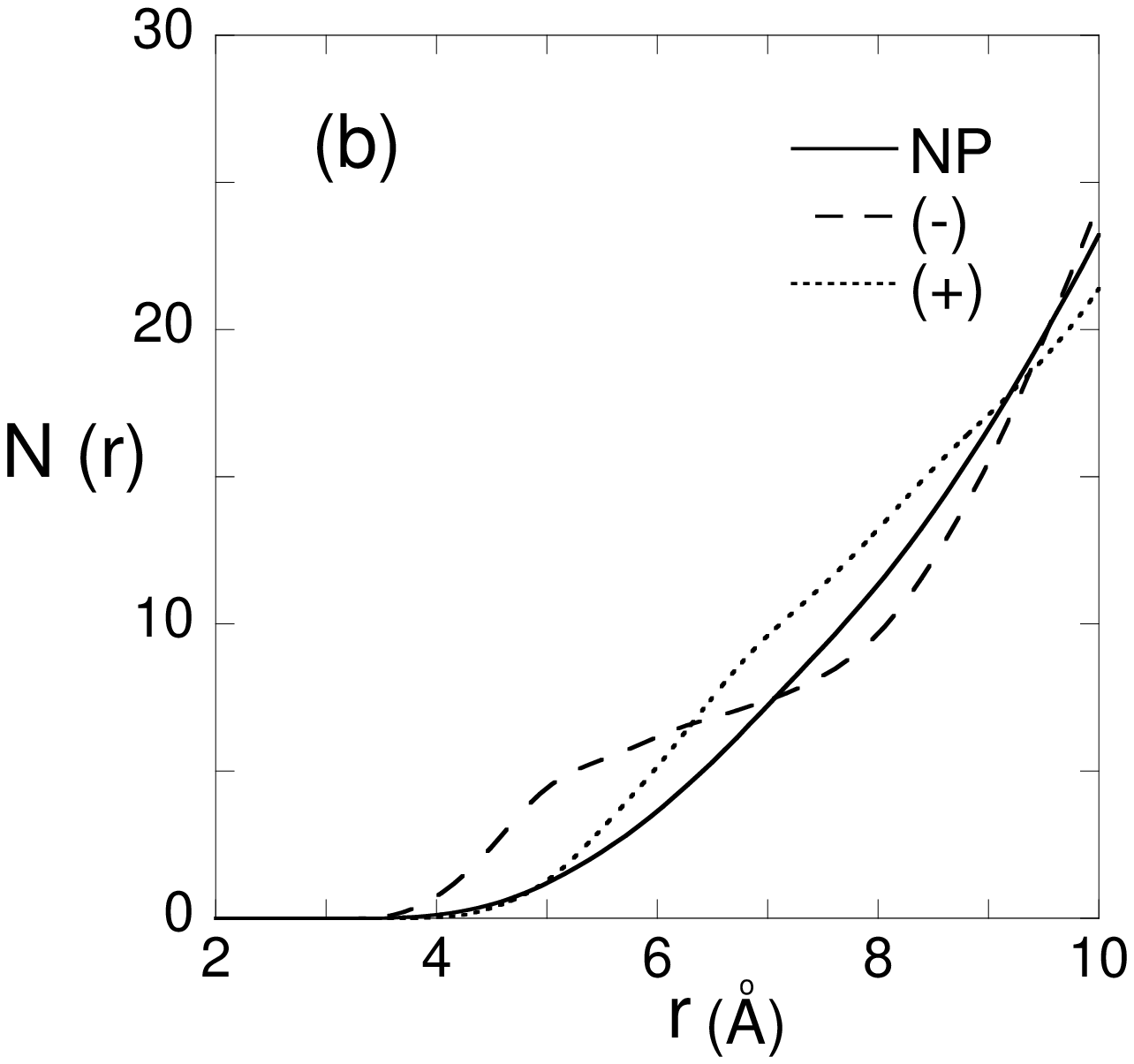}\vspace{0.3in}
\vspace{10cm}
\centerline{Fig.~\ref{fig3ad}}

\newpage

\centering
\includegraphics[width=4.0in]{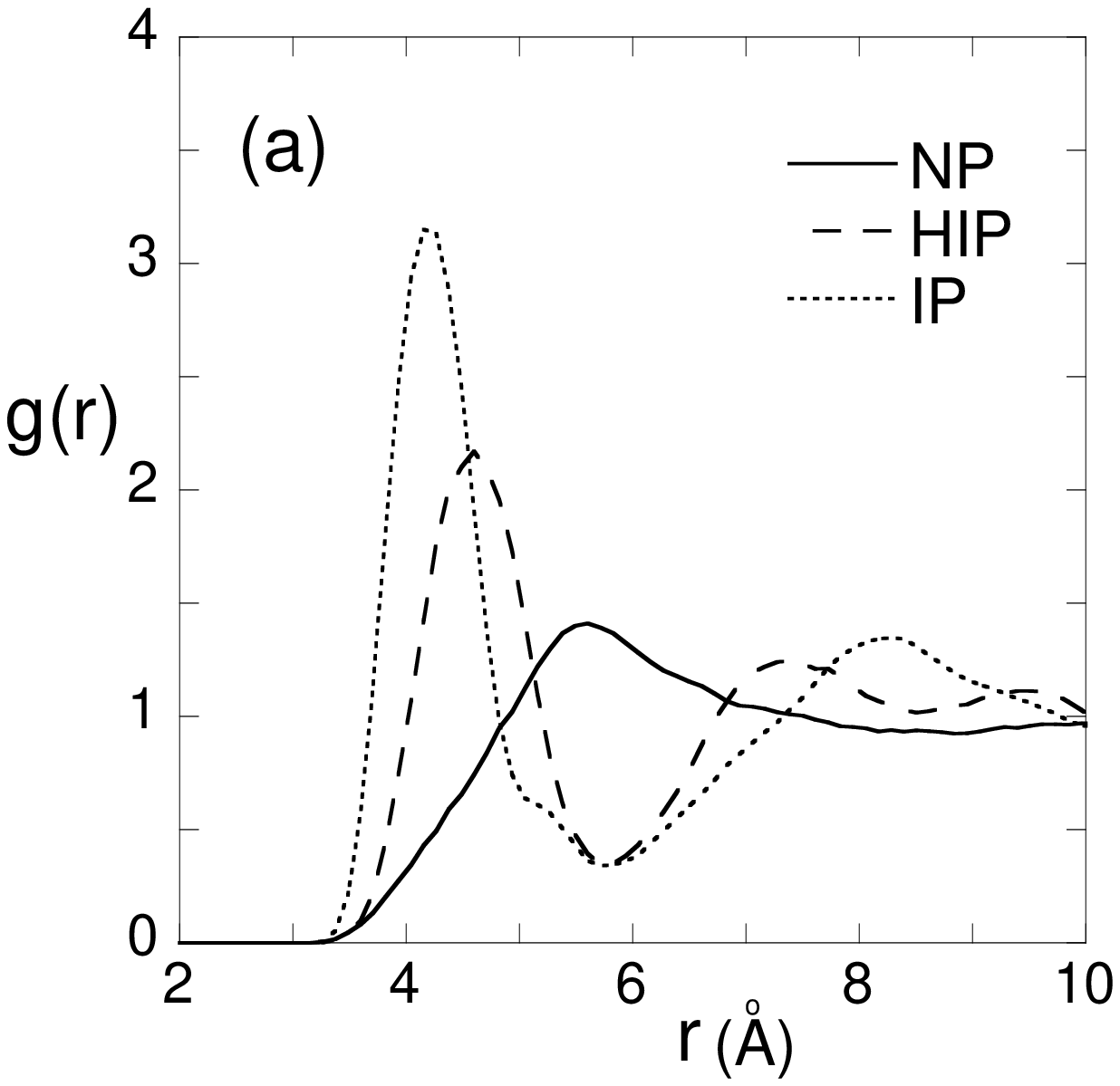}\vspace{0.5in}
\includegraphics[width=4.0in]{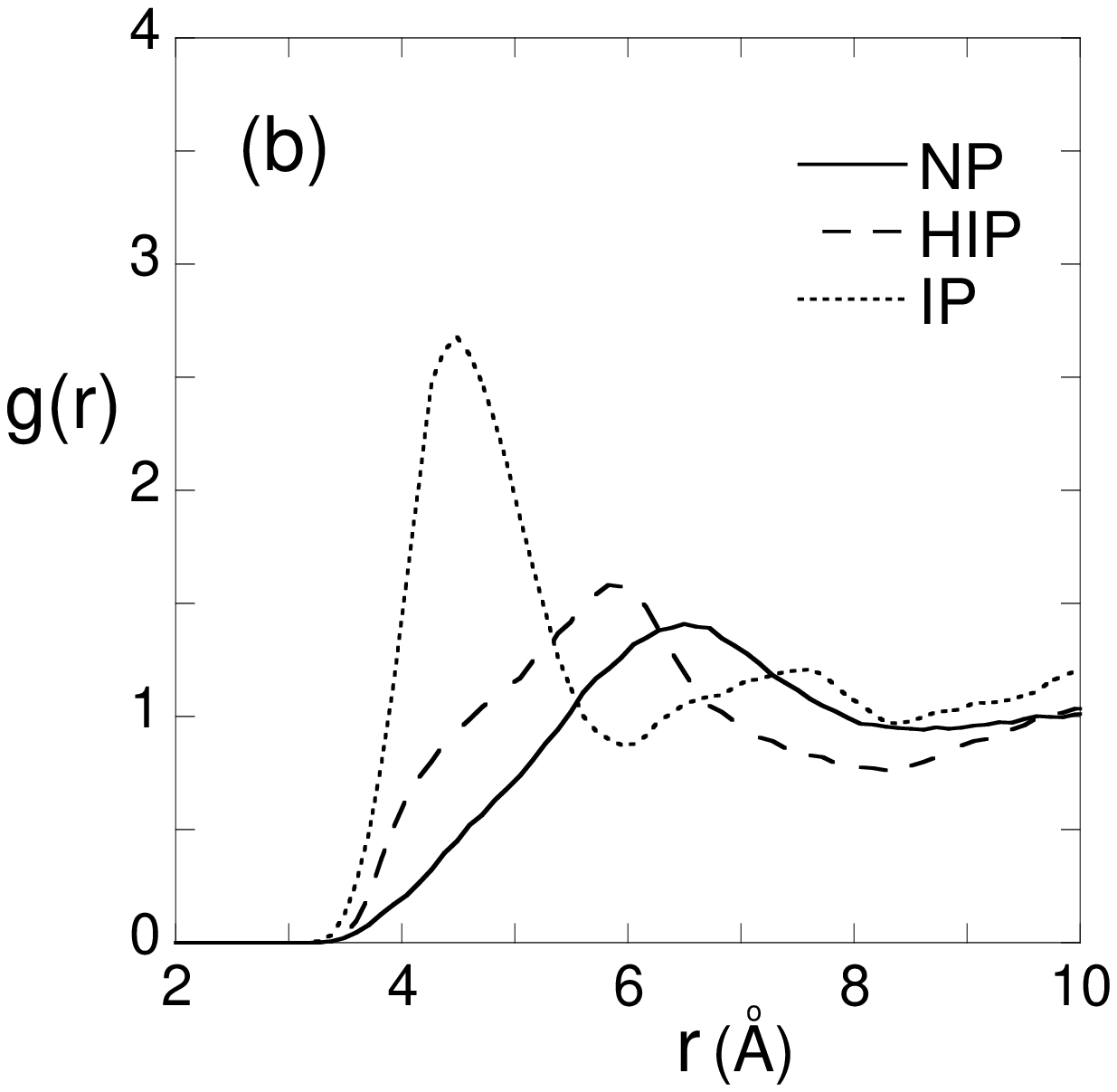}\vspace{0.3in}
\vspace{10cm}
\centerline{Fig.~\ref{fig3}}

\newpage

\centering 
\begin{minipage}{8.0cm} 
\centering 
\includegraphics[width=3.2in]{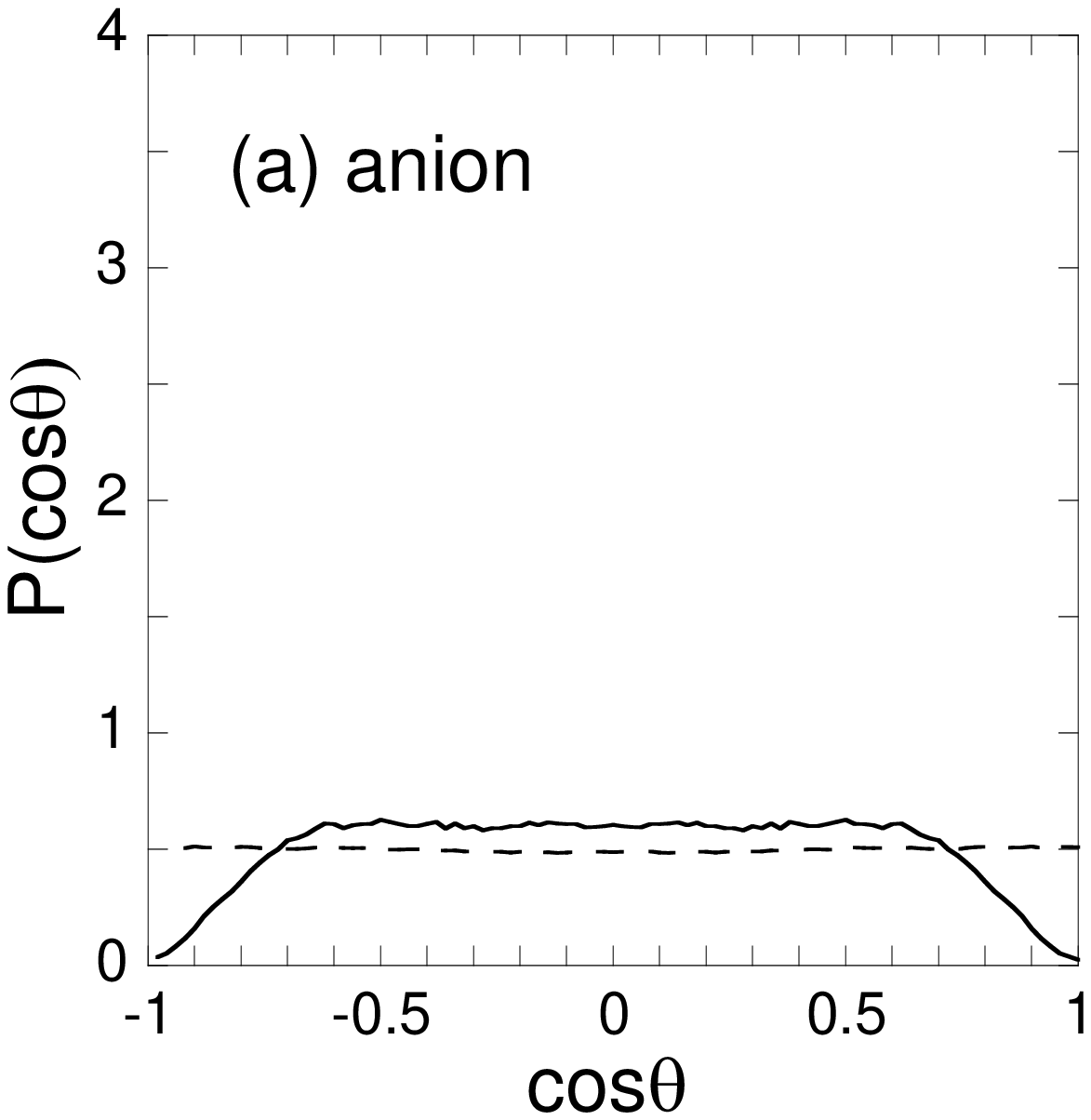} 
\end{minipage} 
\begin{minipage}{8.0cm} 
\centering 
\includegraphics[width=3.2in]{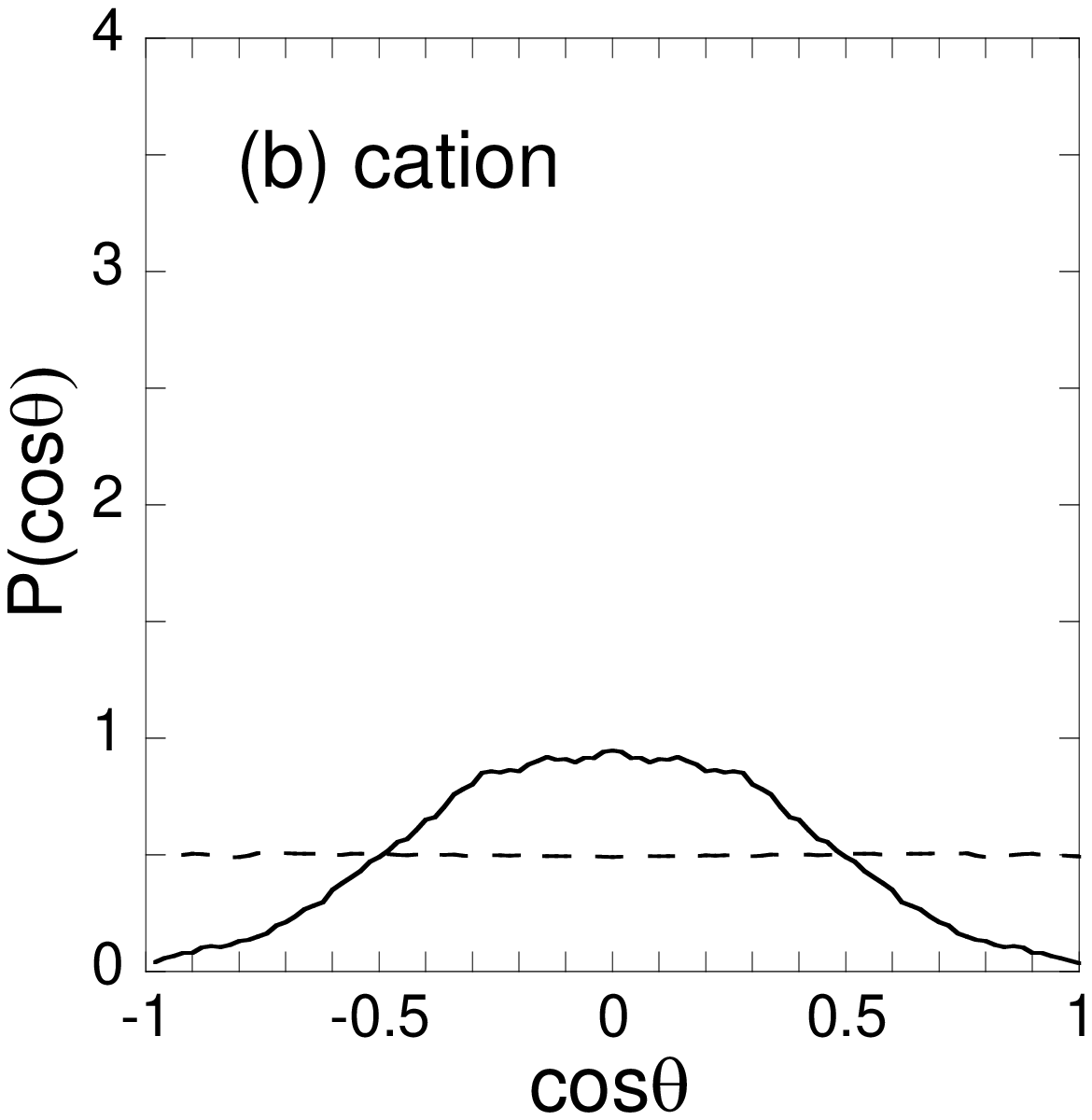} 
\end{minipage} \\[0.5cm] 
\begin{minipage}{8.0cm} 
\centering 
\includegraphics[width=3.2in]{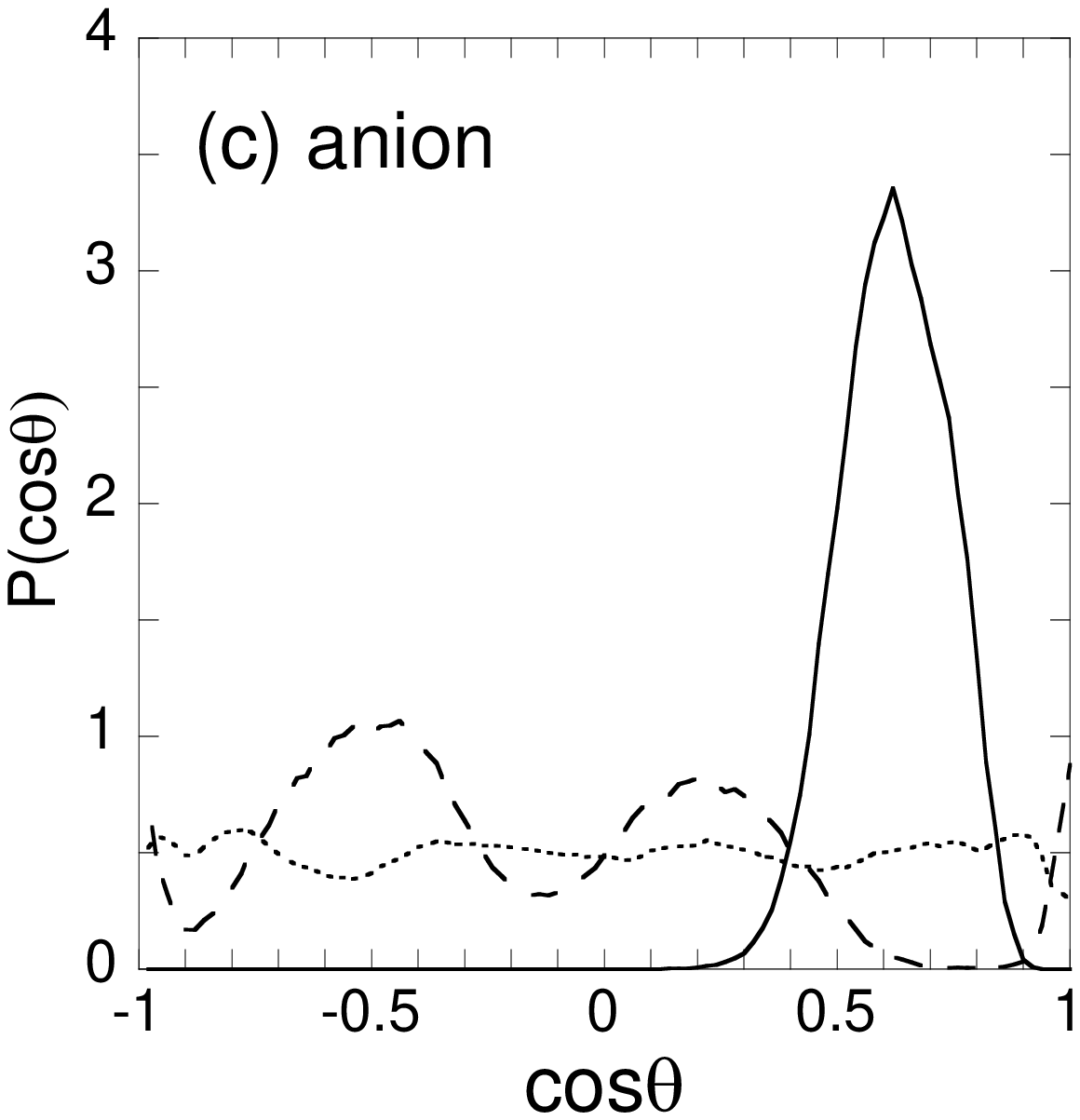} 
\end{minipage} 
\begin{minipage}{8.0cm} 
\centering 
\includegraphics[width=3.2in]{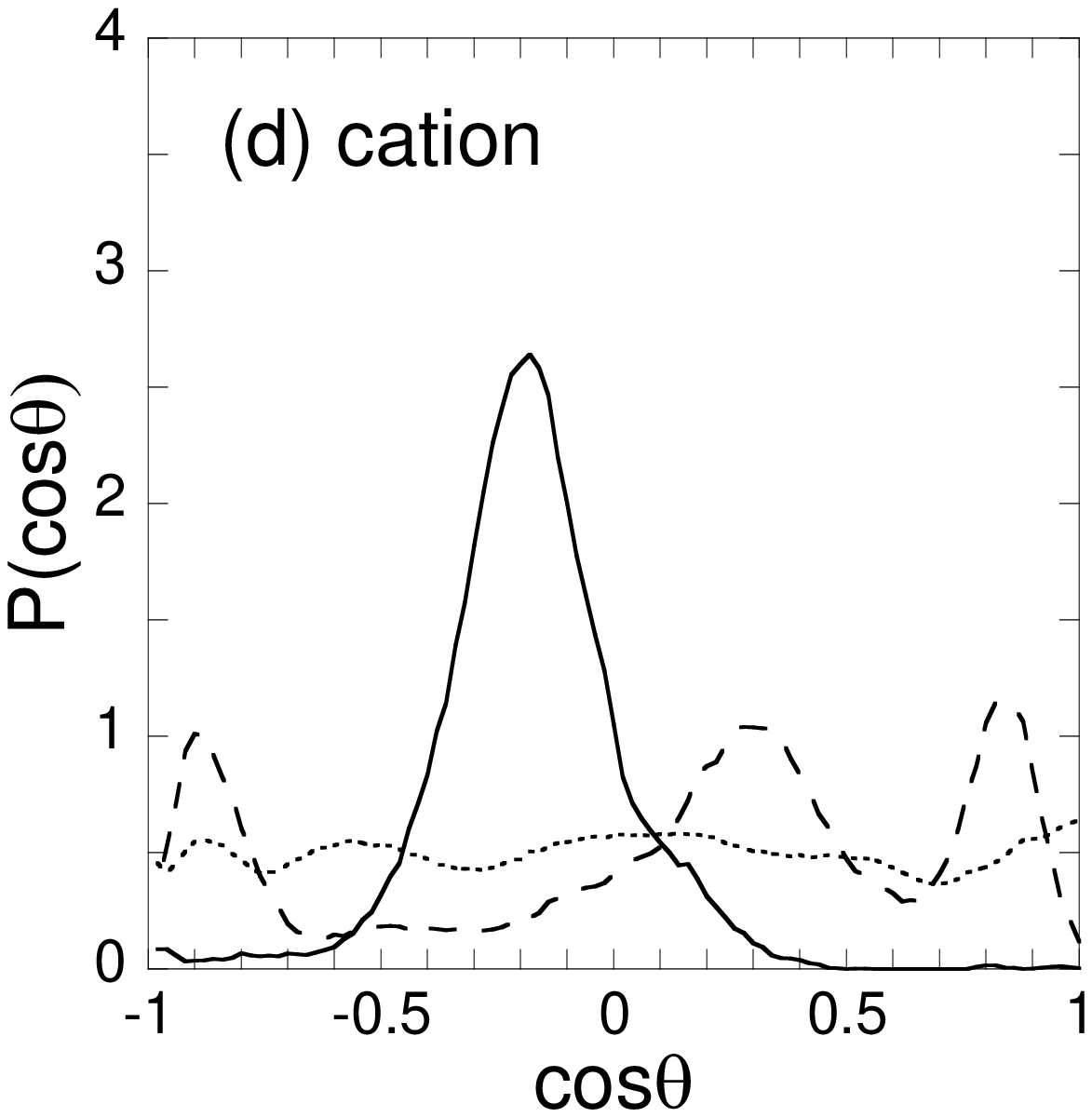} 
\end{minipage} 
\vspace{10cm}
\centerline{Fig.~\ref{fig4}}

\newpage

\centering 
\begin{minipage}{8.0cm} 
\centering 
\includegraphics[width=4in]{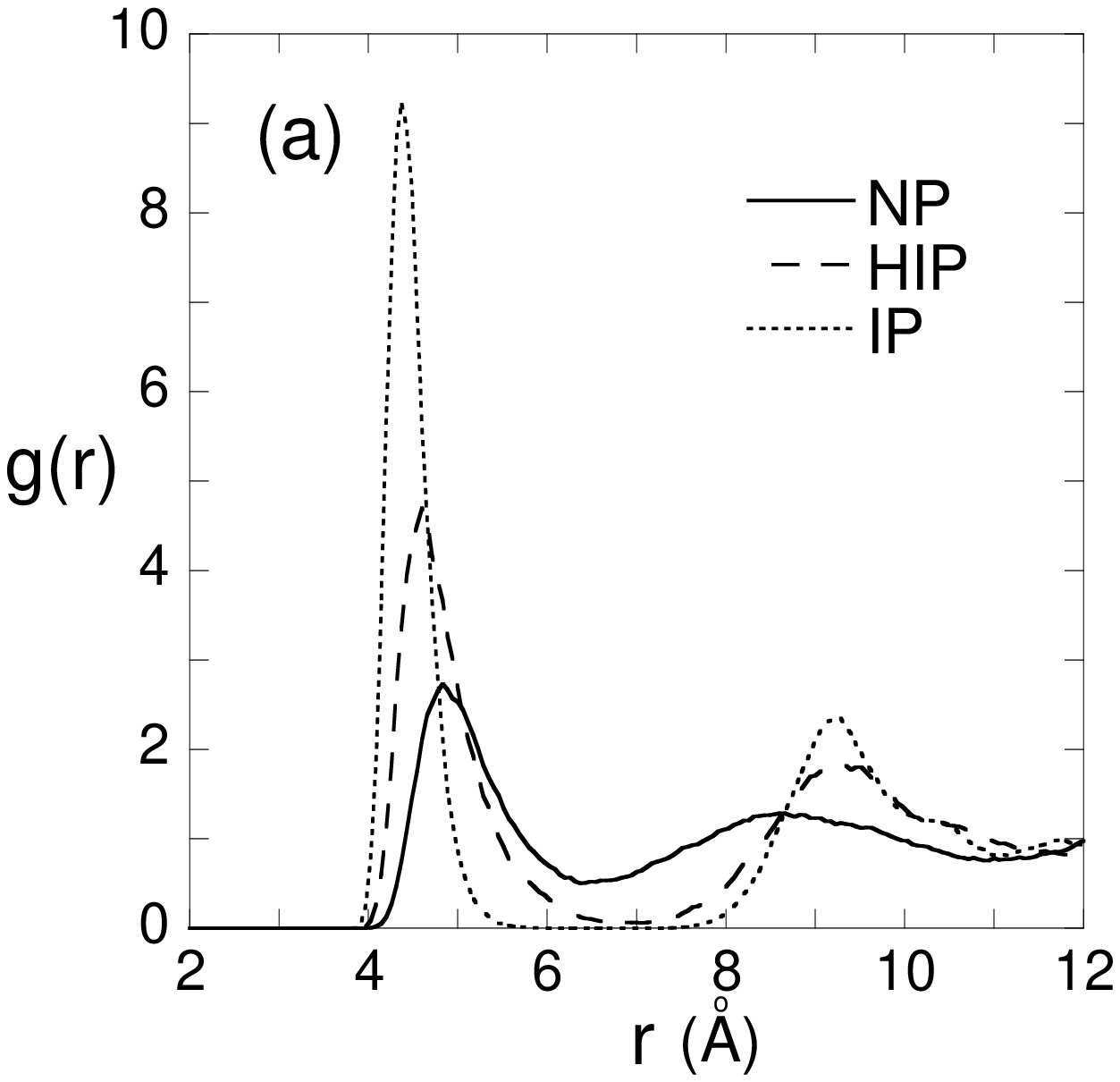} 
\end{minipage} 
\begin{minipage}{8.0cm} 
\centering 
\includegraphics[width=3.6in]{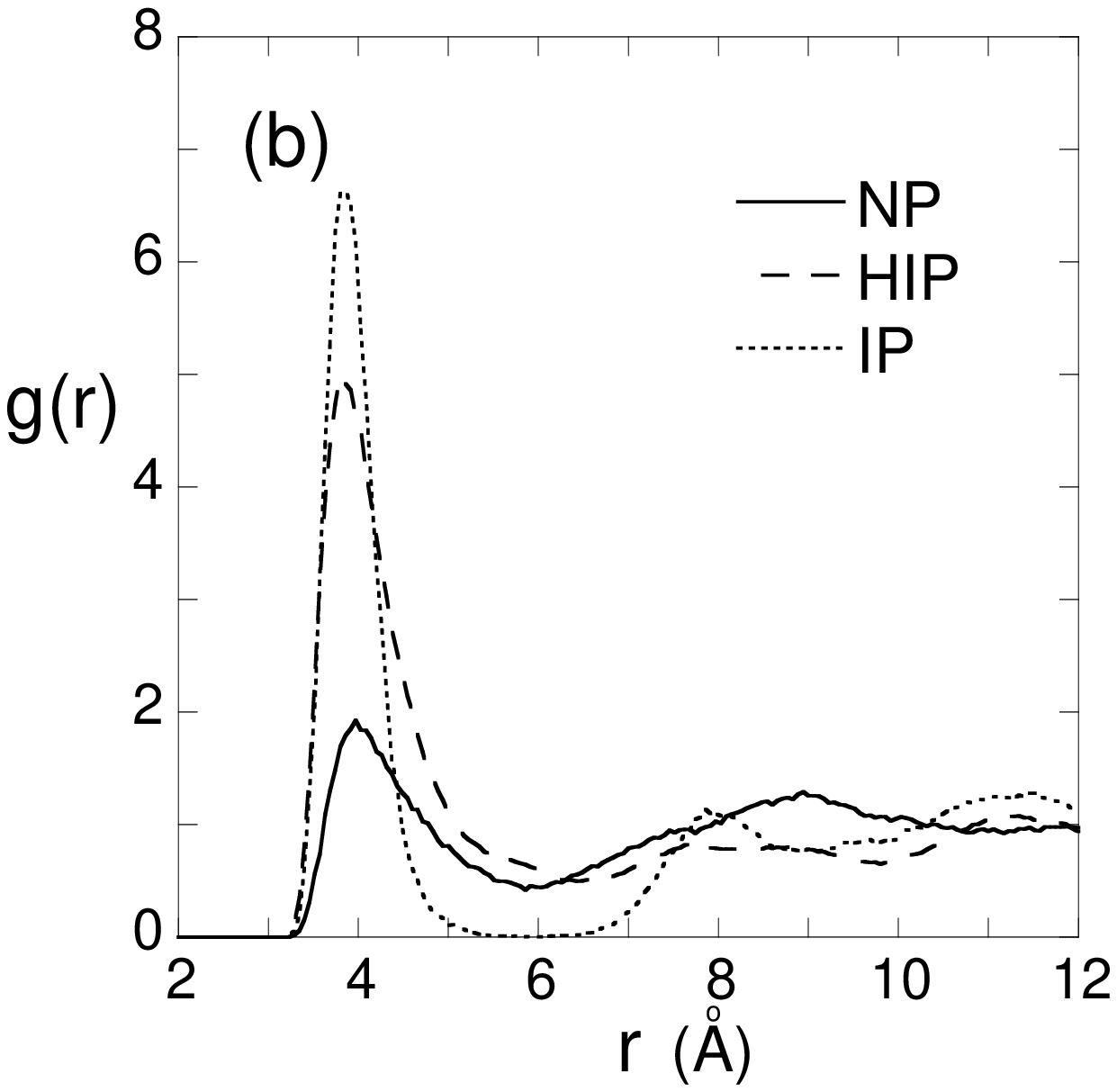} 
\end{minipage} \\[0.5cm] 
\begin{minipage}{8.0cm} 
\centering 
\includegraphics[width=3.6in]{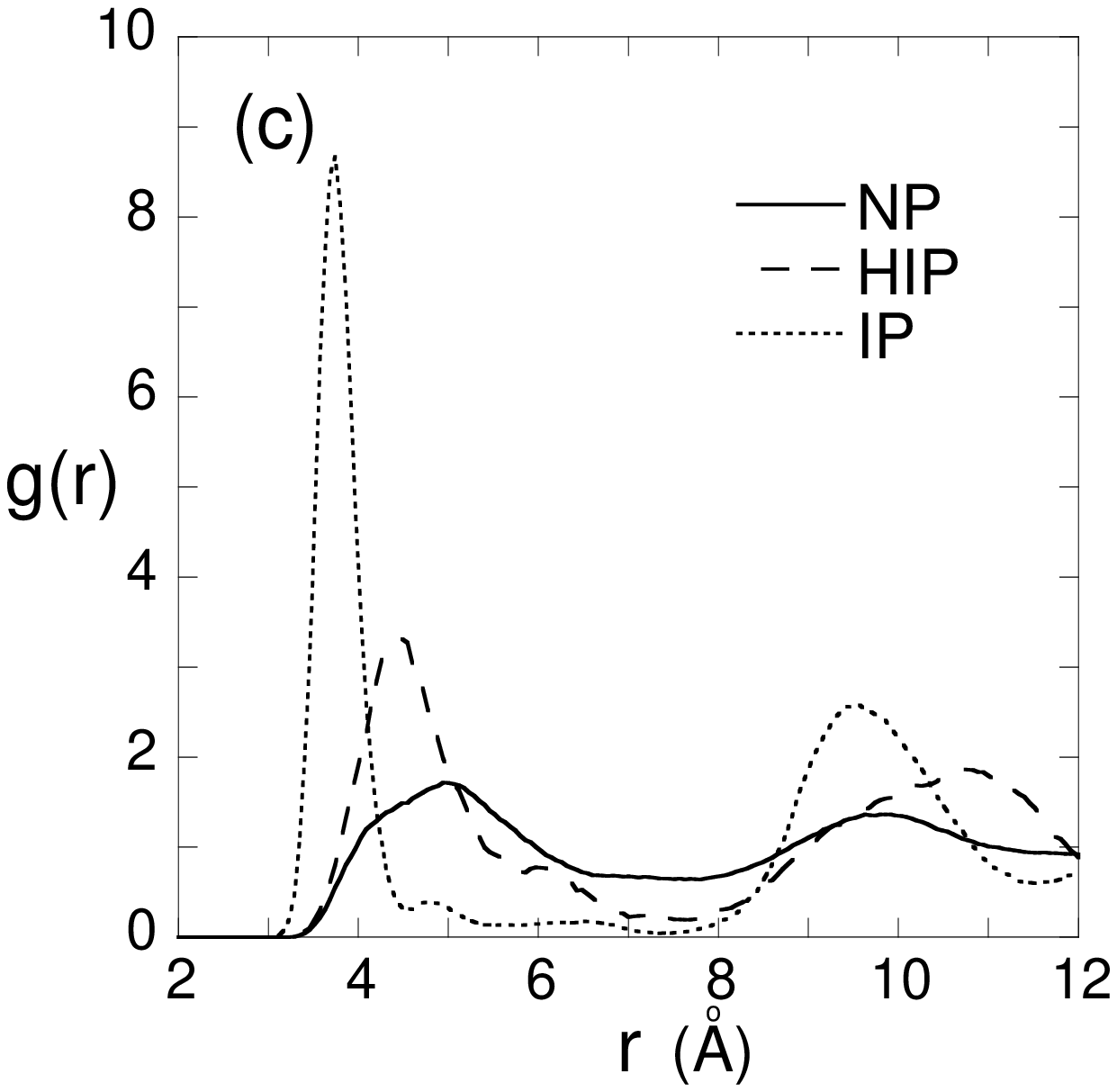} 
\end{minipage} 
\vspace{10cm}
\centerline{Fig.~\ref{fig5}}

\newpage

\centering 
\begin{minipage}{8.0cm} 
\centering 
\includegraphics[width=3.5in]{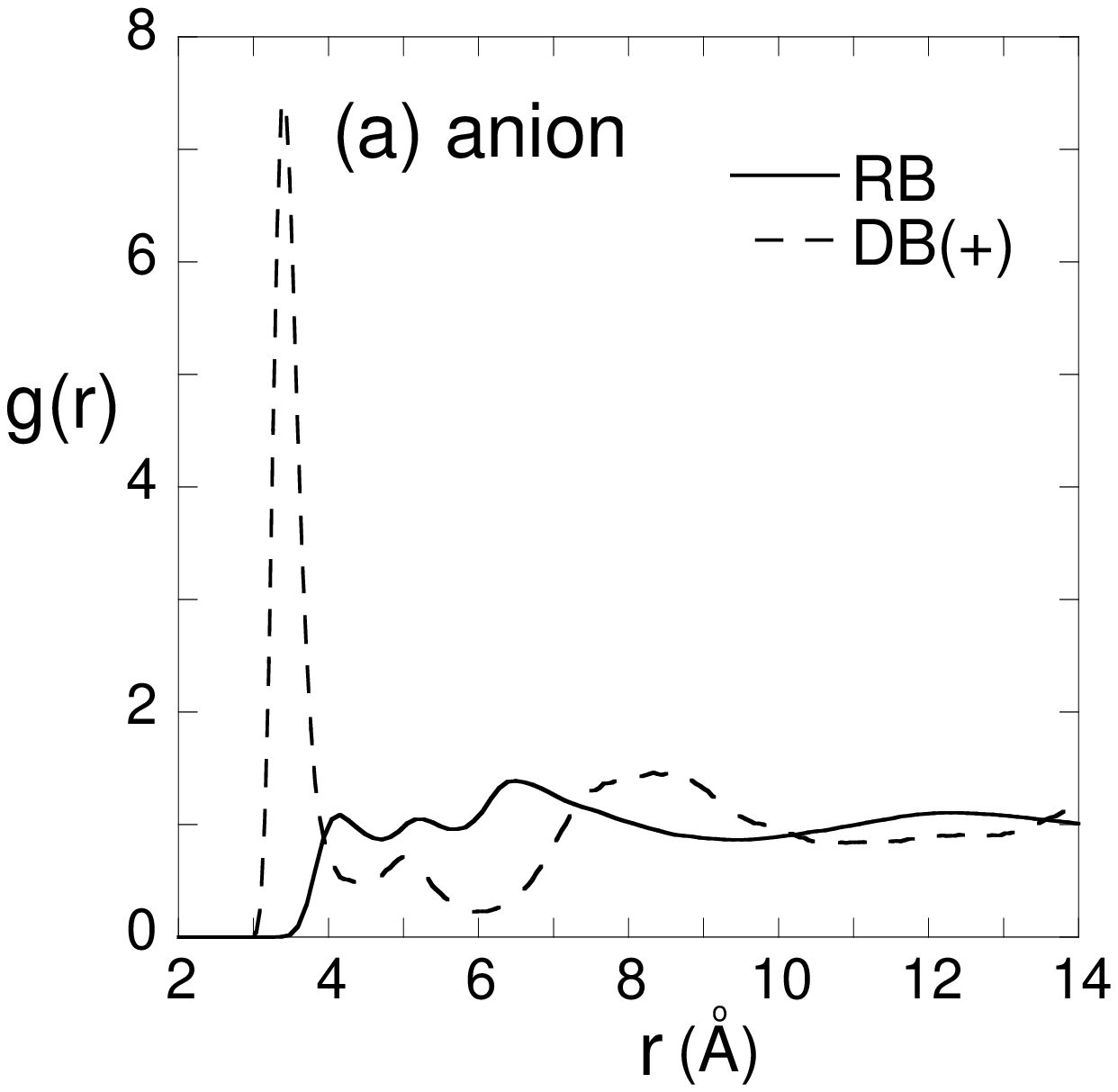} \\  \vspace*{-15pt} 
\end{minipage} 
\begin{minipage}{8.0cm} 
\centering 
\includegraphics[width=3.5in]{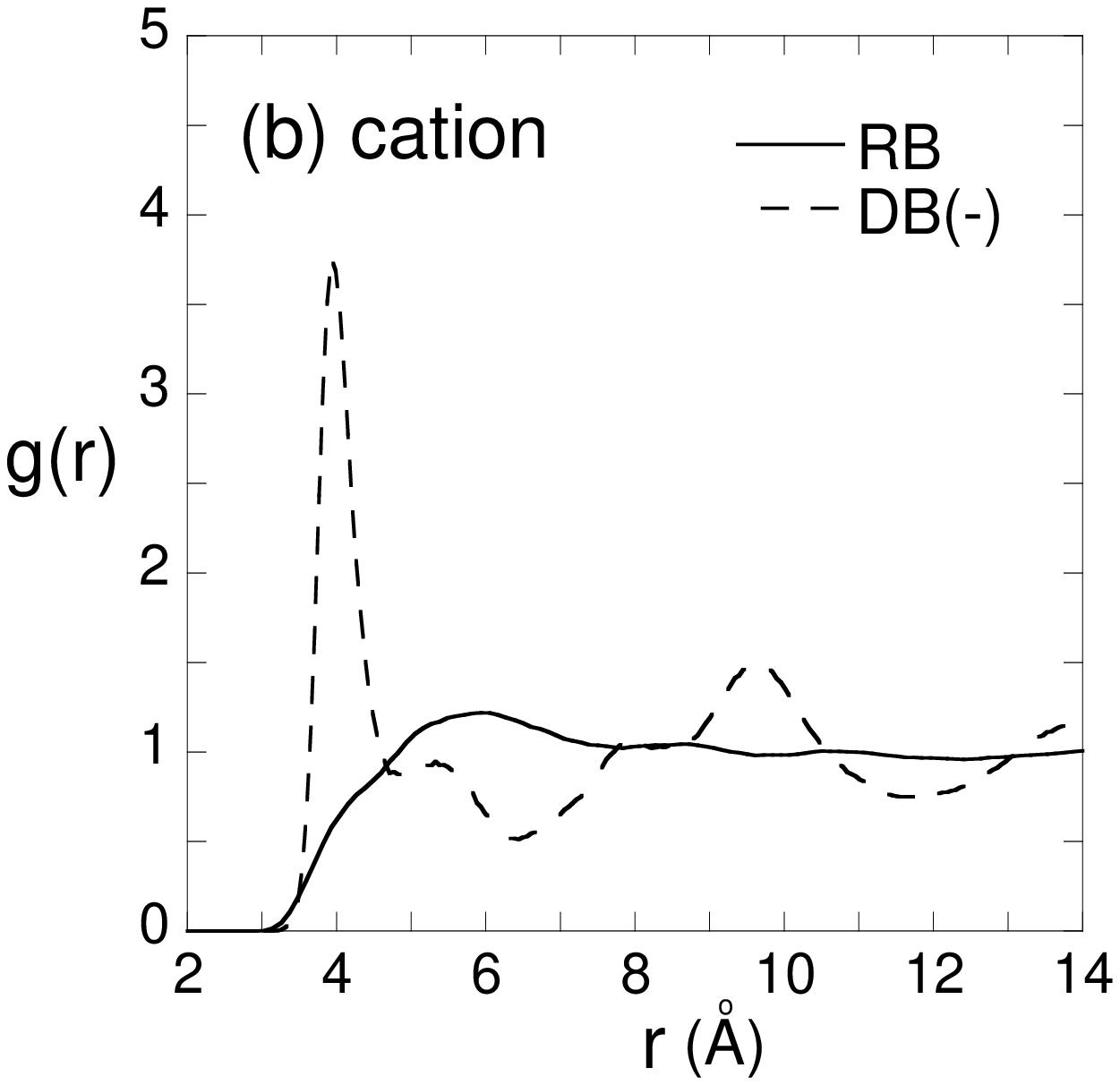} \\  \vspace*{-15pt} 
\end{minipage} \\[0.2in] 
\begin{minipage}{8.0cm} 
\centering 
\includegraphics[width=3.5in]{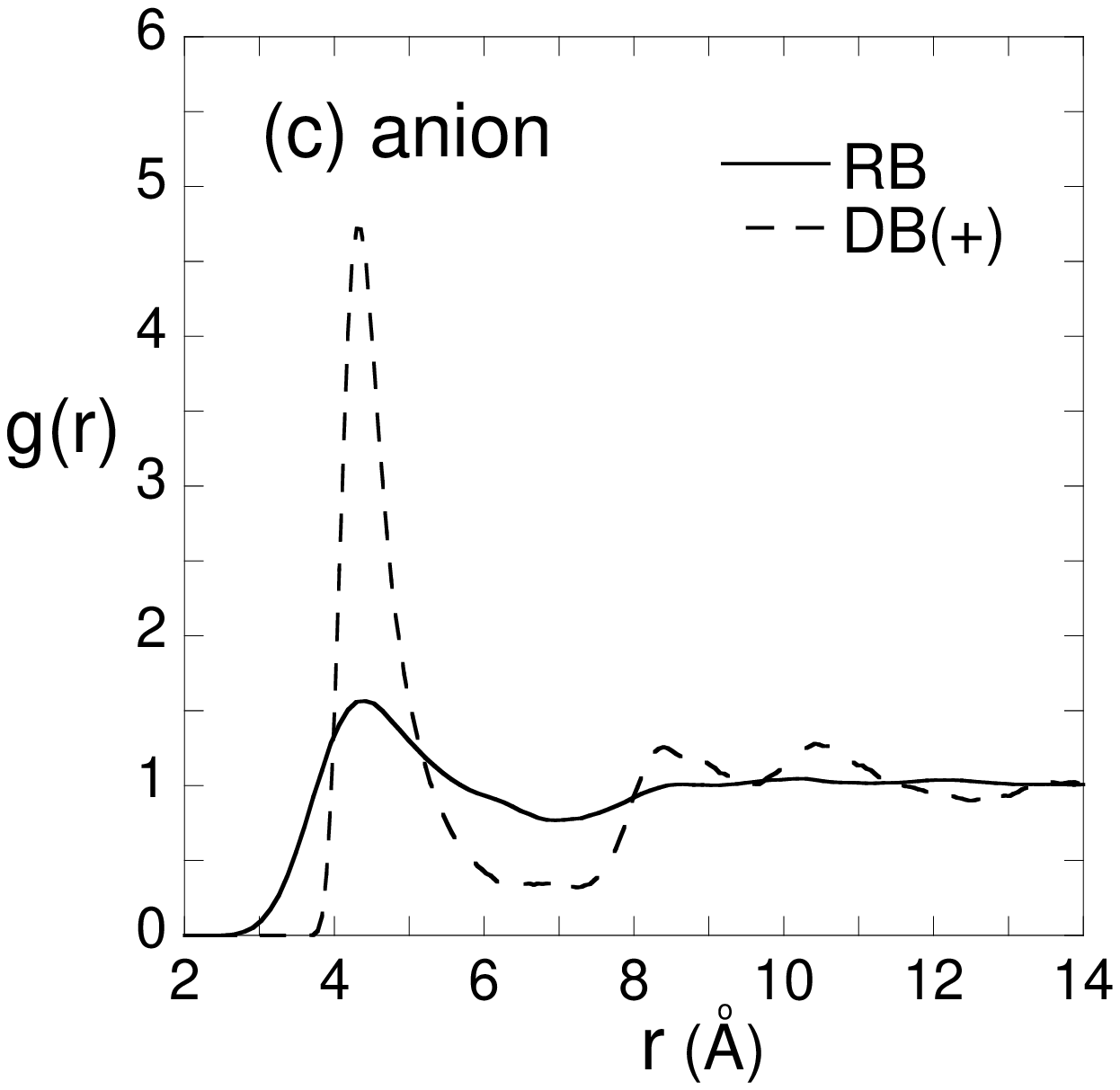} \\  \vspace*{-15pt} 
\end{minipage} 
\begin{minipage}{8.0cm} 
\centering 
\includegraphics[width=3.5in]{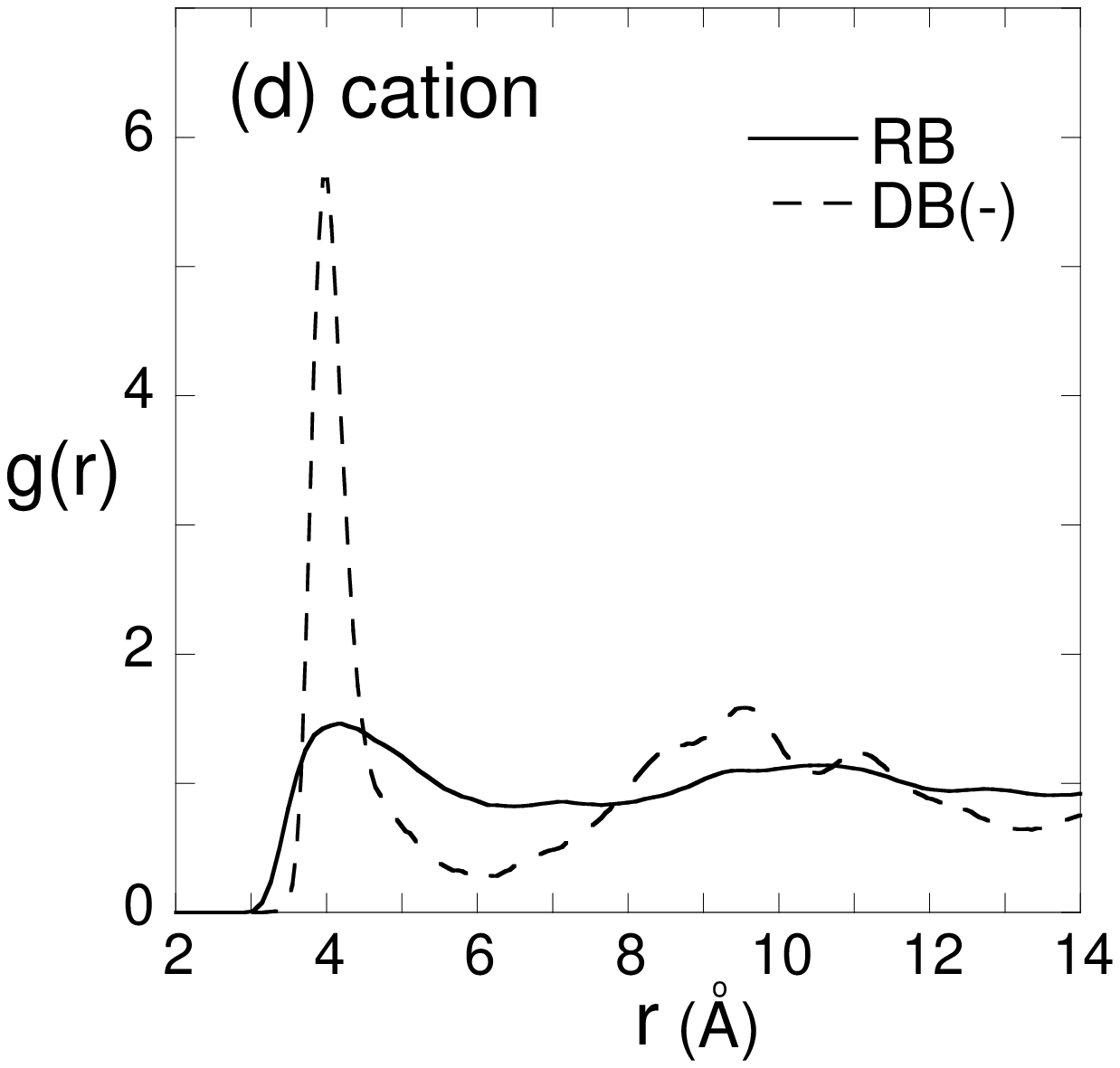} \\  \vspace*{-15pt} 
\end{minipage} 
\vspace{10cm}
\centerline{Fig.~\ref{fig6}}

\newpage

\centering
\includegraphics[width=4.0in]{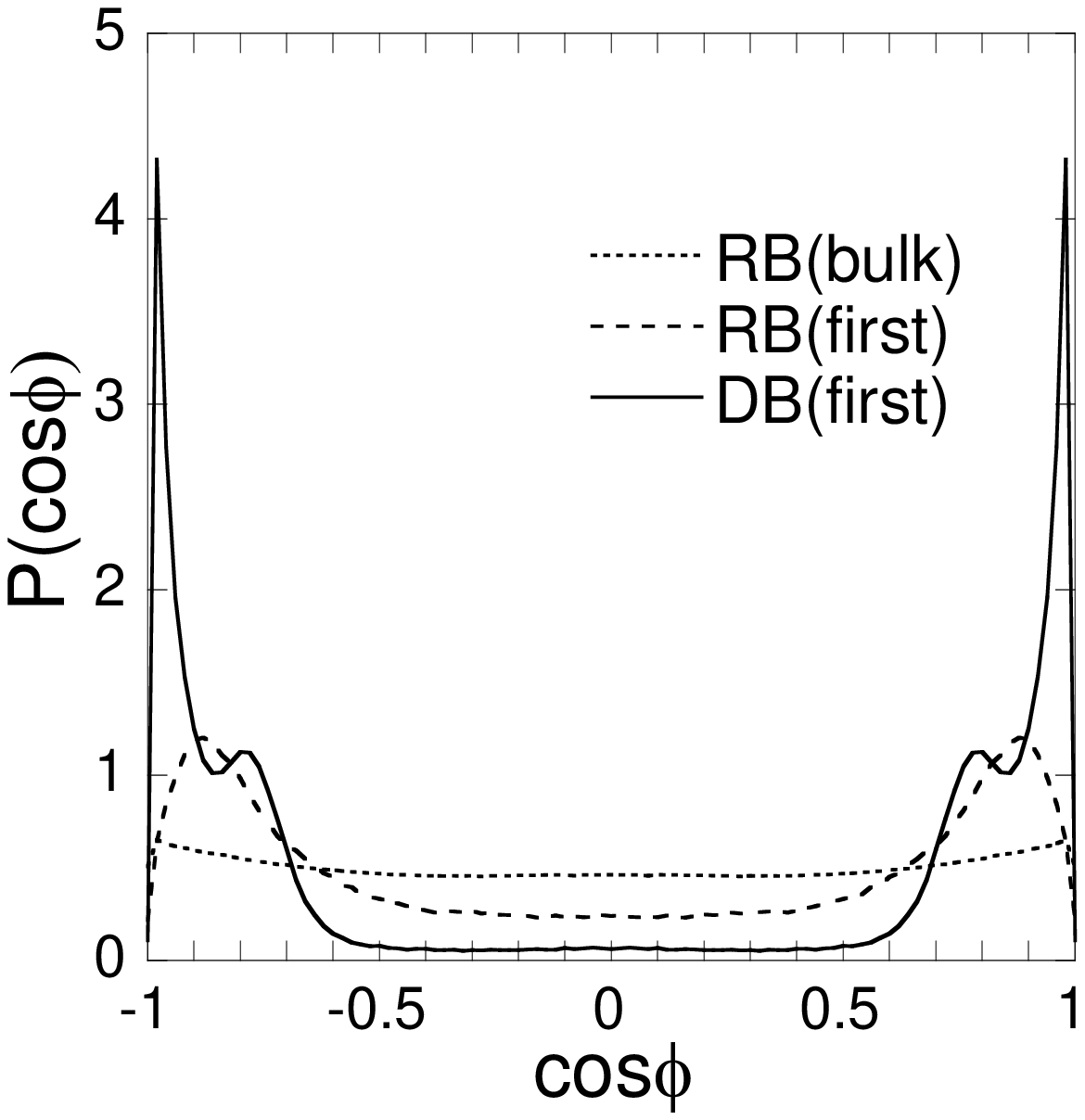}\vspace{0.3in}
\vspace{10cm}
\centerline{Fig.~\ref{fig:ring}}

\newpage

\centering
\includegraphics[width=4.0in]{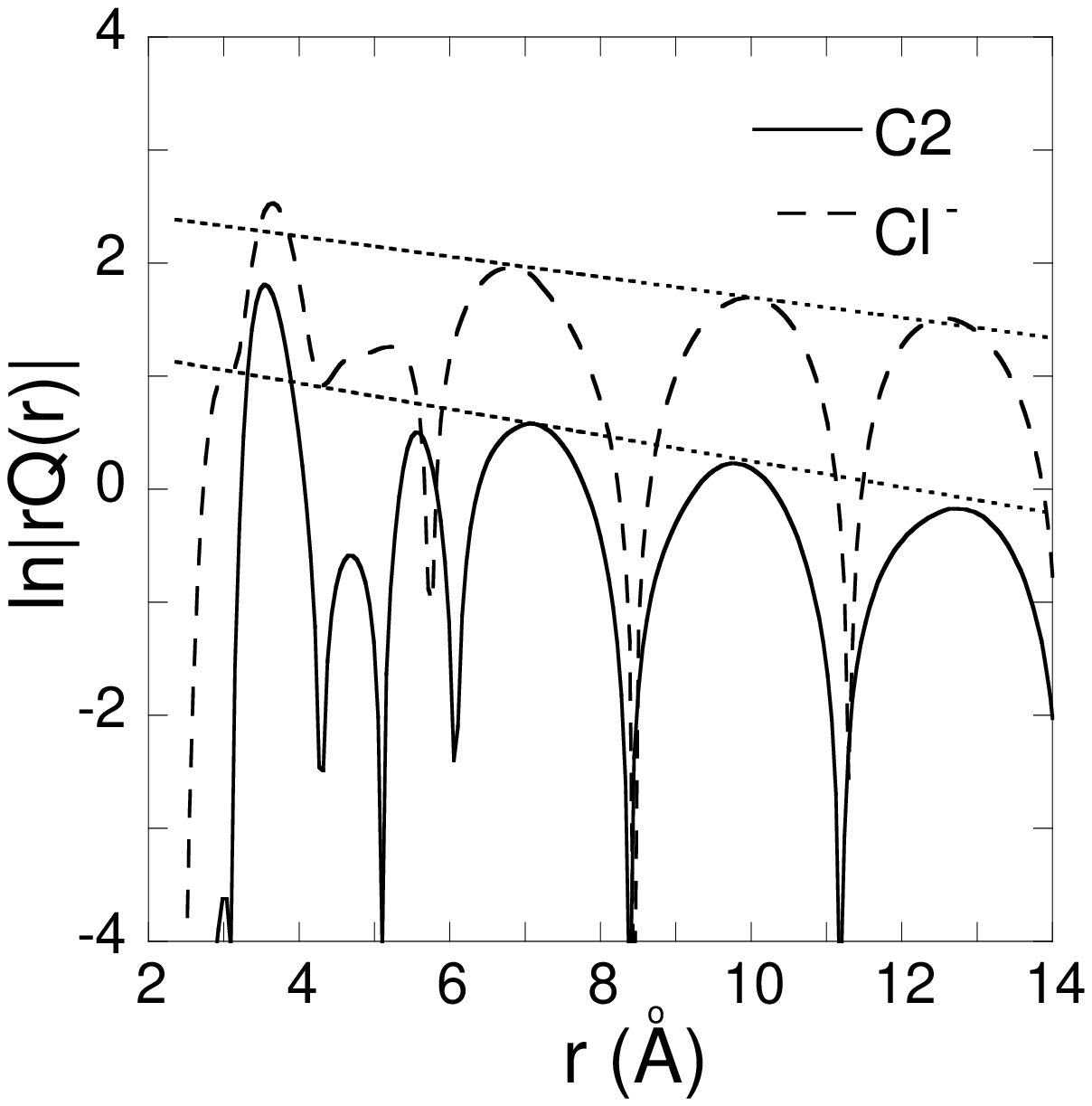}\vspace{0.3in}
\vspace{10cm}
\centerline{Fig.~\ref{fig:screen}}

\newpage

\centering
\includegraphics[width=4.0in]{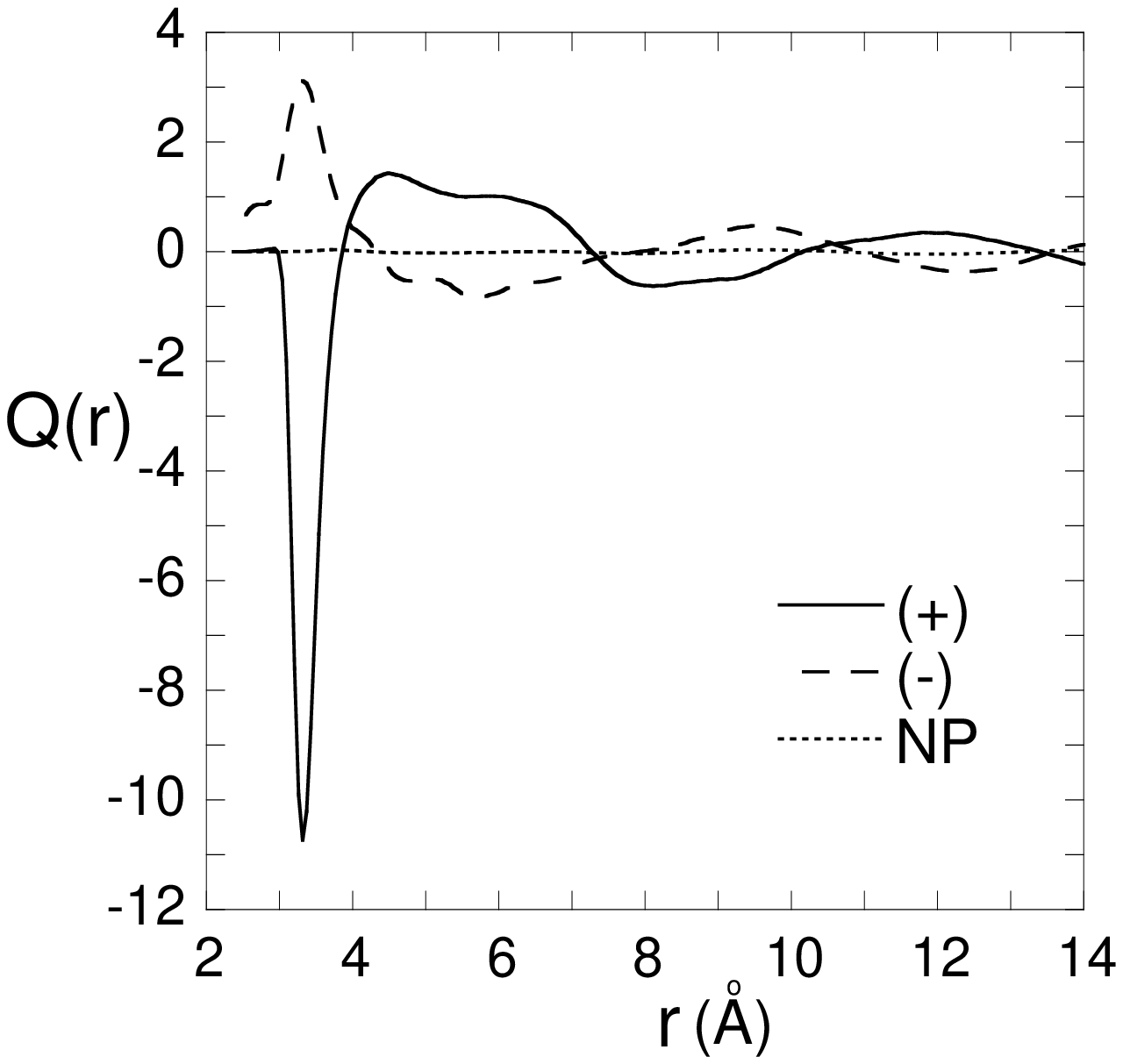}\vspace{0.3in}
\vspace{10cm}
\centerline{Fig.~\ref{fig:screen_DP}}

\newpage

\centering
\includegraphics[width=3.5in]{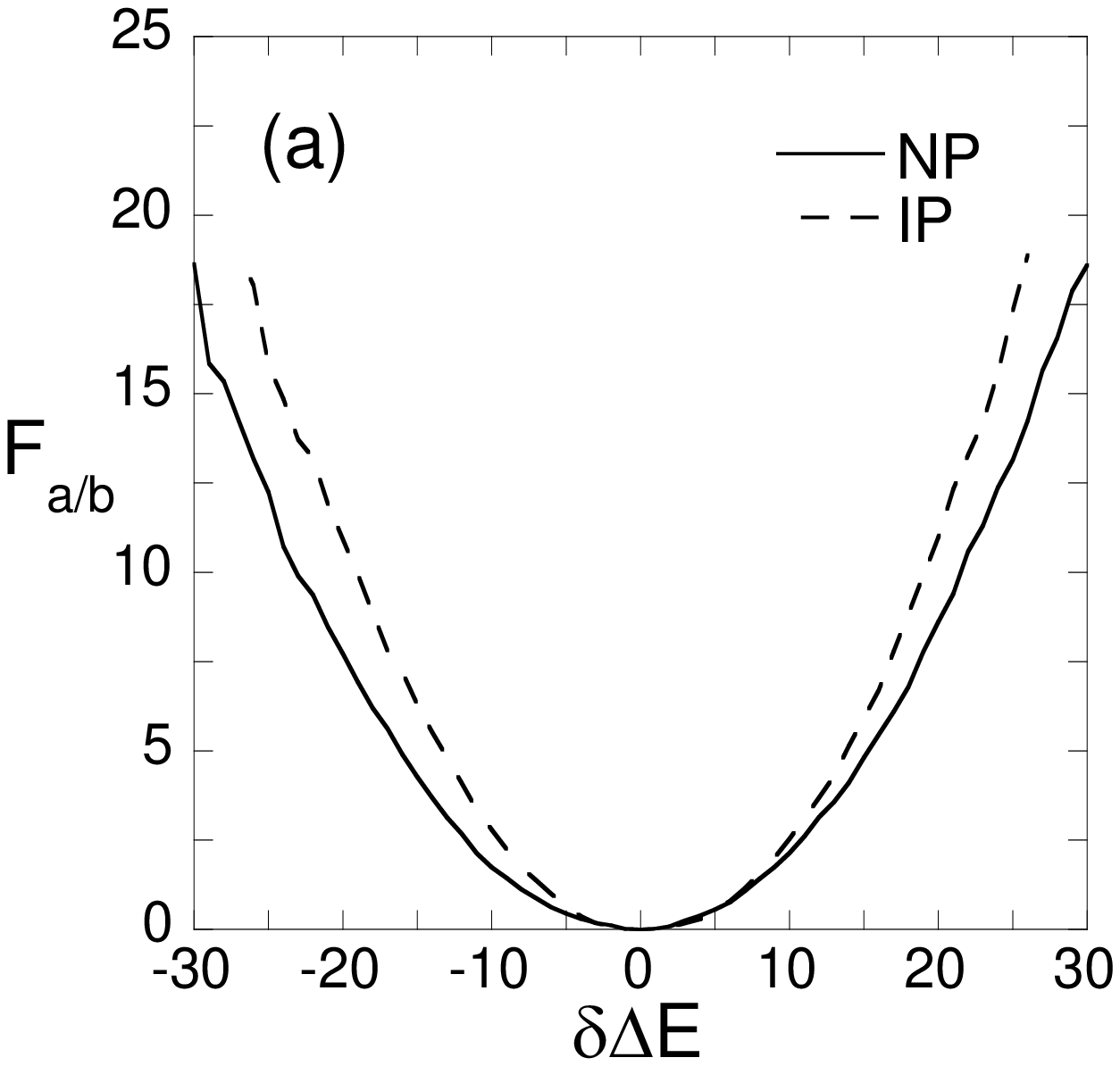}\vspace{0.5in} 
\includegraphics[width=3.5in]{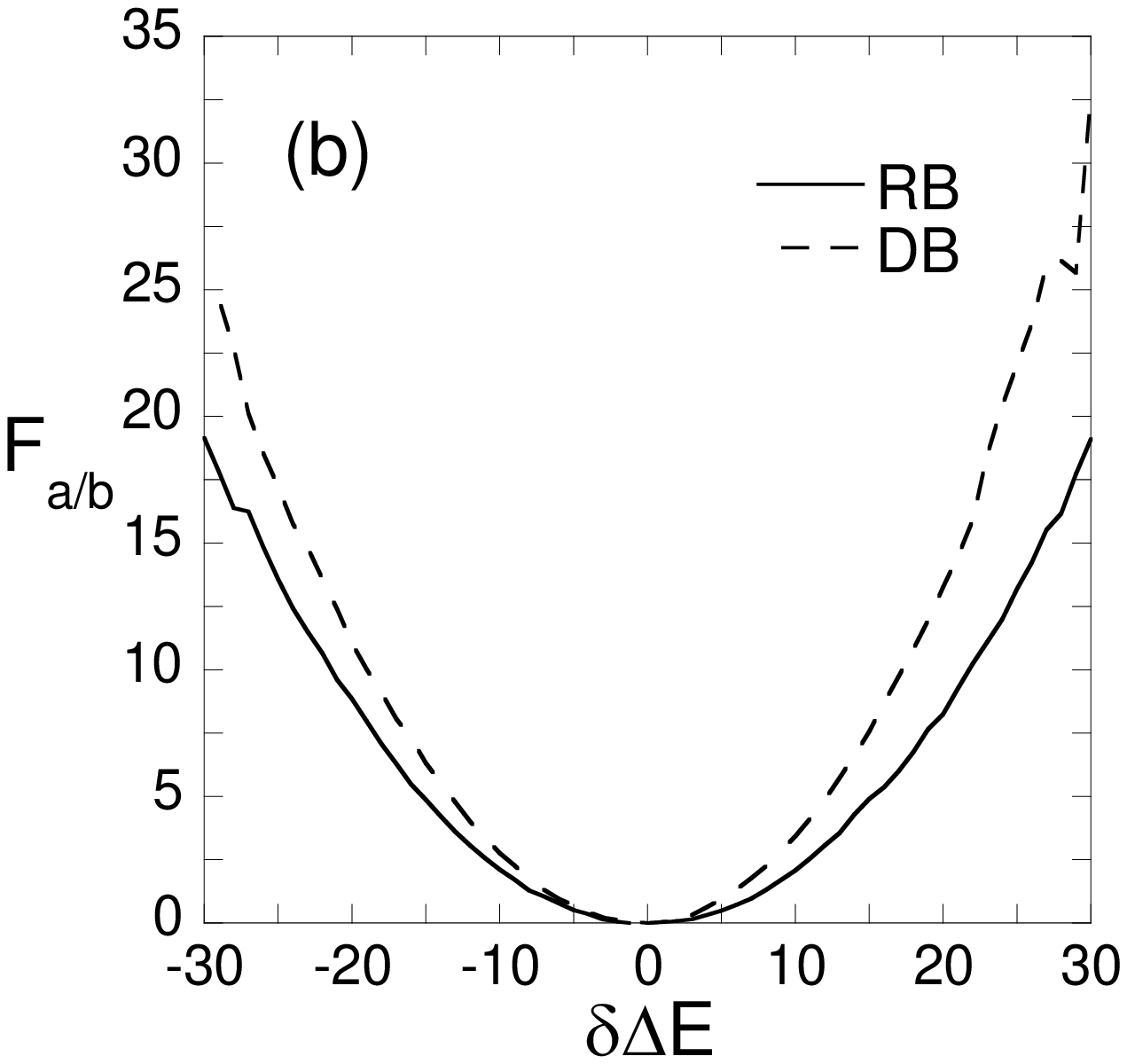}\vspace{0.3in} 
\vspace{10cm}
\centerline{Fig.~\ref{fig:free}}

\end{document}